\documentclass[prd,aps,showpacs,epsf,floats,onecolumn]{revtex4}%
\usepackage{amssymb}
\usepackage{amsfonts}
\usepackage{amsmath}
\usepackage{graphicx}
\usepackage{fancyhdr}%
\providecommand{\U}[1]{\protect\rule{.1in}{.1in}}
\begin{document}
\title{\textbf{Complexity of Pure and Mixed Qubit Geodesic Paths on Curved Manifolds}}
\author{\textbf{Carlo Cafaro}$^{1}$ and \textbf{Paul M.\ Alsing}$^{2}$}
\affiliation{$^{1}$SUNY Polytechnic Institute, 12203 Albany, New York, USA}
\affiliation{$^{2}$Air Force Research Laboratory, Information Directorate, 13441 Rome, New
York, USA}

\begin{abstract}
It is known that mixed quantum states are highly entropic states of imperfect
knowledge (i.e., incomplete information) about a quantum system, while pure
quantum states are states of perfect knowledge (i.e., complete information)
with vanishing von Neumann entropy. In this paper, we propose an information
geometric theoretical construct to describe and, to a certain extent,
understand the complex behavior of evolutions of quantum systems in pure and
mixed states. The comparative analysis is probabilistic in nature, it uses a
complexity measure that relies on a temporal averaging procedure along with a
long-time limit, and is limited to analyzing expected geodesic evolutions on
the underlying manifolds. More specifically, we study the complexity of
geodesic paths on the manifolds of single-qubit pure and mixed quantum states
equipped with the Fubini-Study metric and the Sj\"{o}qvist metric,
respectively. We analytically show that the evolution of mixed quantum states
in the Bloch ball is more\ complex than the evolution of pure states on the
Bloch sphere. We also verify that the ranking based on our proposed measure of
complexity, a quantity that represents the asymptotic temporal behavior of an
averaged volume of the region explored on the manifold during the evolution of
the systems, agrees with the geodesic length-based ranking. Finally, focusing
on geodesic lengths and curvature properties in manifolds of mixed quantum
states, we observed a softening of the complexity on the Bures manifold
compared to the Sj\"{o}qvist manifold.

\end{abstract}

\maketitle

%======== fancyhdr: puts the page number back in on the [R]ight vs the [L]eft =====
\fancyhead[L]{\ifnum\value{page}<2\relax\else\thepage\fi}
%================================================================

%=================
%needed for fancyhdr
%=================
\thispagestyle{fancy}
%===============

\pagebreak

\section{Introduction}

We divide the Introduction in three parts to better motivate the selection of
our goals along with their physical relevance. In the first part, we highlight
the use of geometric concepts originally introduced in quantum computing and
later borrowed by high energy physicists to describe and, to a certain extent,
understand black holes behavior. In particular, we emphasize the geometric
characterization of some complexity notions, including gate complexity and
state complexity. In the second part, we outline several distinguishing
features that characterize the physics of systems specified by pure and mixed
quantum states. Neither geometry nor complexity are mentioned in this second
part. In the third part, we finally describe our main objectives.

\subsection{Geometry in quantum computing and high energy physics}

Geometry plays a fundamental role in science \cite{pettini00,pettini07},
including quantum computing and high energy physics. In Ref. \cite{nielsen06},
Nielsen and collaborators used methods of Riemannian geometry to propose a way
of finding efficient quantum circuits capable of performing certain
computational tasks. They proposed a geometric measure of quantum algorithm
complexity for quantum circuits constructed with unitary gates. Their
formalism led to a geometric continuous-time version of the discrete gate
complexity, a measure of complexity quantifying how hard it is to build a
unitary operator \cite{nielsen10,2020}. In such geometric context, finding
optimal quantum circuits is equivalent to finding the shortest path between
two points in a certain curved geometry. Essentially, one introduces a
Riemannian metric in the space of unitary operators acting on a given number
of qubits. The metric quantifies how hard it is to implement a given quantum
computational task. Then, the distance induced by the metric in the space of
unitary operators is employed as a measure of the complexity of the quantum
operation. In addition to gate complexity, one can also introduce in quantum
information science the concept of quantum computational complexity of a
state, a measure quantifying how hard it is to build a unitary transformation
that transforms the reference state to the target state \cite{nielsen10}.
Geometric concepts (including actions, path lengths, volumes, and complexity)
play a fundamental role in high energy physics as well. For instance, quantum
computational complexity measures of geometric origin appear to play a
fundamental role in encoding properties of the interiors of black holes
\cite{leo18}. In Refs. \cite{leo14,leo15}, it was shown that the quantum
computational complexity of the dual quantum state is proportional to the
spatial volume of the Einstein-Rosen bridge (i.e., a structure linking two
sides of the Penrose diagram of an eternal anti-de Sitter black hole). In
Refs. \cite{leo16,leo16B}, it was argued that the quantum computational
complexity of a holographic state is proportional to the action of a certain
spacetime region termed Wheeler-DeWitt patch. For very insightful applications
of Nielsen's geometric approach to quantum computational complexity of states
and gates in the single-qubit and multi-qubit scenarios of special relevance
in high energy physics, we refer to Refs. \cite{brown19} and \cite{auzzi21},
respectively. The analysis in Ref. \cite{brown19} is rather illuminating
because it clearly shows the effects of replacing a non deformed Bloch sphere
equipped with the usual Fubini-Study metric with a deformed Bloch sphere with
a new metric that does not treat all directions in the tangent space in a
similar manner. Indeed, in Nielsen's geometric approach, the single-qubit
Hilbert space is equipped with a metric that stretches directions that are
hard to move in, assigning them a large distance. Two main consequences of
this new metric can be summarized as follows: First, geodesics are no longer
generated by time-independent Hamiltonians. Second, suitable choices of the
anisotropy penalty factors specifying the new metric can lead to spaces with
negative sectional curvature. This, in turn, is responsible of the chaotic
growth of perturbations (i.e., exponential maximal complexity). For a work
focusing on the connection between a geometric measure of quantum
computational complexity and negative curvature, we refer to Ref.
\cite{brown17}. The work in Ref. \cite{auzzi21} attracts great interest for
several reasons, including the fact that it addresses the issue of ergodicity
of geodesics on manifolds of negative curvature. This is especially important
in view of a potential application of thermodynamical arguments to complexity
evolution. As previously pointed out, Nielsen's approach to quantum
computation defines a geometric measure on the space of unitary operators. In
Ref. \cite{chapman18}, instead, the Fubini-Study metric is used to define a
geometry on the space of states to propose a complexity measure assigned to a
target state. This complexity is the minimal distance as measured by the
Fubini-Study metric among all parametrized curves on the space of states that
connect the reference state to the desired target state. Within this approach,
the Fubini-Study metric accounts for the complexity by keeping track of the
changes of the state (by means of applications of unitary operations)
throughout the preparation of the target state. In Ref. \cite{ruan21}, a
notion of mixed state complexity is extended to impure quantum states by
replacing the Fubini-Study metric with the Bures metric (or, alternatively,
the quantum Fisher information metric) and, at the same time, extending the
nature of quantum transformations acting on the state to non-unitary
operations. Finally, following what happens for pure states, the complexity of
mixed states is identified with the (Bures) length of the geodesic connecting
the reference and target mixed states. To a certain extent and to the best of
our knowledge, given the novelty of the introduction of the concept of mixed
state complexity, no comparative analysis exists in the literature between
complexity behaviors associated to physical systems specified by pure and
mixed states. We intend to cover this point in this paper.\begin{table}[t]
\centering
\begin{tabular}
[c]{c|c|c|c|c|c}\hline\hline
\textbf{Knowledge of system} & \textbf{Type of state} & \textbf{Purity} &
\textbf{Von Neumann entropy} & \textbf{Temperature} & \textbf{Entanglement}%
\\\hline
Complete & Pure & Maximal & Minimal & Low & Typical\\
Partial & Mixed & Not maximal & Not minimal & High & Less typical\\\hline
\end{tabular}
\caption{Schematic description of physical systems in pure and mixed quantum
states in terms of purity, von Neumann entropy, temperature, and
entanglement.}%
\end{table}

\subsection{Pure and mixed quantum states}

In quantum information science, when one has complete knowledge about a
quantum system, one can use a pure state to describe it. However, complete
knowledge is only available in limiting ideal (noiseless) scenarios (i.e.,
isolated/closed quantum systems). In practice, one only has partial knowledge
about a quantum system. Indeed, small errors may happen in the preparation,
evolution, or measurement of the system due to imperfect devices or to
(external) coupling with other degrees of freedom outside of the system that
one is controlling. In these realistic (noisy) cases (i.e., open quantum
systems), quantum systems are described by mixed states. These states are
specified by classical probability distributions over pure states and are used
to represent our probabilistic ignorance of a pure state. The density operator
formalism is a very powerful mathematical tool for incorporating a lack of
complete knowledge about a quantum system. Within this formalism, the
\textquotedblleft quantumness\textquotedblright\ of the system resides in the
off-diagonal entries of the density matrix. These are interference terms
between the pure states that specify the mixture that defines the mixed state.
A particular measure of noisiness of a quantum state is the purity
\textrm{P}$\left(  \rho\right)  \overset{\text{def}}{=}\mathrm{Tr}\left(
\rho^{2}\right)  $ of a density operator $\rho$. The purity of a pure state is
equal to one, and the purity of a mixed state is strictly less than one with
$1/N\leq$\textrm{P}$\left(  \rho\right)  \leq1$ for an $N\times N$ density
matrix. The departure of a system from a pure state can also be quantified by
means of the von Neumann entropy \textrm{S}$_{vN}\left(  \rho\right)
\overset{\text{def}}{=}-\mathrm{Tr}\left(  \rho\log\rho\right)  $. This
quantity specifies the degree of mixing of the state describing a given
finite-dimensional quantum system. For a pure state, the van Neumann entropy
vanishes. Instead, for a maximally mixed state characterized by the complete
absence of off-diagonal entries in the density matrix (thus, describing
something non-interfering and seemingly classical), the von Neumann is maximal
and equals $\log\left(  2\right)  $ for a qubit system.

To the best of our knowledge, there does not exist any comparative geometric
analysis of the complexity of pure and mixed states in the literature. From an
intuitive standpoint, there are several reasons why one expects mixed states
to be more complex than pure states: 1) Mixed states are generally used to
describe highly entropic systems that can exhibit a temperature higher than
the one specifying systems in a pure state. In statistical mechanics, for
instance, a physical system at thermal equilibrium is described by a thermal
(Gibbs) state \cite{huang}. The Gibbs state is a mixed state with a
well-defined finite temperature value. However, at zero temperature (i.e.,
$\beta\overset{\text{def}}{=}(k_{B}T)^{-1}\rightarrow\infty$ with $k_{B}$
denoting the Boltzmann constant), the system is in a pure state. In this
limiting case, the density matrix has every element zero except for a single
element on the diagonal. At infinite temperature (i.e., $\beta\rightarrow0$),
instead, the system is in a maximally mixed state (i.e., a mixture of pure
states with equal statistical weights). For example, consider a spin-$1/2$
particle in a stationary and uniform magnetic field $B_{0}$ along the
$z$-direction. The Hamiltonian of the system can be written as \textrm{H}%
$\overset{\text{def}}{=}(\hslash\omega_{0}/2)\sigma_{z}$ with $\omega
_{0}\overset{\text{def}}{=}\left(  eB_{0}\right)  /m$. Clearly, $e$ and $m$
denote the electric charge and the mass of the electron, respectively.
Moreover, $\hslash\overset{\text{def}}{=}h/(2\pi)$ is the reduced Planck
constant and, finally, $\sigma_{z}$ is the Pauli phase flip operator. At
thermal equilibrium, the density matrix of the system is given by
$\rho_{\mathrm{TE}}\left(  \beta\right)  \overset{\text{def}}{=}%
e^{-\beta\mathrm{H}}/\mathrm{Tr}(e^{-\beta\mathrm{H}})$.\ A simple calculation
shows that $\rho_{\mathrm{TE}}\left(  \beta\right)  $ becomes a maximally
mixed (or, pure) state as $T$ approaches infinity (or, zero). For a definition
of temperature of arbitrary quantum states, beyond thermal (Gibbs) states used
for physical systems at thermal equilibrium, we refer to Ref. \cite{brunner22}%
. The temperature quantifies the ability of a quantum system to cool down or
heat up a thermal environment in Ref. \cite{brunner22}. Finally, for a scheme
to measure the temperature of individual pure quantum states by means of
quantum interference, we refer to Ref. \cite{mitch22}; 2) Systems in mixed
states are less quantum (or, alternatively, more classical) than systems in
pure states. In Ref. \cite{karol98}, it is proven that entanglement, a
quintessential quantum property of physical systems, is typical of pure
states, while separability is connected with quantum mixtures. For intriguing
connections of geometric flavor among purity, separability, and complex
behavior in quantum scattering processes, we refer to Refs. \cite{kim11,kim12}%
. In stead, for possible justifications of why chaoticity viewed as temporal
complexity is softer in quantum systems compared with classical systems, we
hint at Refs. \cite{caron01,caron04,kroger06,adom12a,adom12b}. Furthermore, it
is known that the existence of speed limits is not something peculiar to
quantum systems \cite{taddei13,campo13,deffner13}. Indeed, there are speed
limits for classical systems as well \cite{oku18,campo18,kosinski21}. In Ref.
\cite{kosinski21}, it was shown that the quantum counterpart of the classical
speed limits derived in Ref. \cite{oku18} are obtained by quantum systems
specified by density operators describing states that become more and more
mixed as $\hslash$ approaches zero; 3) Mixed quantum states can undergo a
richer variety of transformations compared to pure states \cite{nielsen2000}.
In open system dynamics, one needs to consider general nonunitary quantum
evolutions and have the freedom to choose a variety of distance measures
between quantum states. Decoherence and measurements are examples of
noncontrollable and controllable nonunitary processes, respectively. Quantum
channels, for instance, provide us with a formalism for discussing
decoherence, the nonunitary evolution of pure states into mixed states
\cite{stefano12}. In conventional formulations of quantum mechanics, instead,
pure states can only be connected in a unitary fashion. Moreover, the choice
of geometric distance measures between pure states is more restrained than
that between impure states; 4) Mixed states evolutions can exhibit higher
speed values than the ones of pure state temporal changes. In Ref.
\cite{carlini08}, it is shown that the time optimal mixed state evolution can
be faster than the time optimal pure state evolution. In Ref. \cite{deffner13}%
, it is demonstrated that non-Markovian (i.e., memory) effects can speed up
nonunitary quantum evolutions of arbitrarily driven open quantum systems. In
Ref. \cite{girolami19}, it is pointed out that finding the optimal unitary for
mixed target states is more challenging than for pure target states; 5) Mixed
qubit states have three local degrees of freedom, while pure qubit states only
have two local degrees of freedom. From a pure geometric perspective, it is
reasonable to expect that mixed states are more complex than pure states
\cite{hornedal22}. For instance, unlike what happens in optimal-speed unitary
evolutions of systems in pure states, tight evolutions of closed quantum
systems in mixed states are typically generated by time-varying Hamiltonians
\cite{hornedal22}. We refer to Table I for a schematic description of physical
systems in pure and mixed quantum states in terms of purity, von Neumann
entropy, temperature, and entanglement.

Given the lack of a geometric comparative analysis between the complex
behaviors exhibited by physical quantum systems specified by pure and mixed
quantum states and, in addition, given the variety of distinguishing physical
features that characterize the evolution of systems specified by pure and
mixed quantum states, we intend to capture here the complexity of these
evolutions from a geometric standpoint and provide a geometrical picture of
these physical differences.

\subsection{Our goals}

In this paper, we aim to provide a comparative information geometric analysis
of the complexity of geodesic paths of pure and mixed quantum states on the
Bloch sphere and in the Bloch ball, respectively. Our investigation is
partially inspired by the above mentioned geometrically flavored
investigations. Furthermore, it is motivated by our curiosity concerning the
possibility of describing and, to a certain extent, understanding from a
geometric viewpoint the previously mentioned fingerprints of a greater degree
of complexity of mixed quantum states. Finally, it relies on our insights into
the concepts of complexity \cite{cafaro18}, geometric formulations of
optimal-speed Hamiltonian evolutions on the Bloch sphere
\cite{cafaropra20,cafaropra22}, and the role played by the thermodynamic
length and divergence (or, alternatively, action) in studying the complexity
of minimum entropy production probability paths in quantum mechanical
evolutions \cite{cafaropre20,cafaropre22}. The main questions that we address
in this paper can be summarized as follows:

\begin{enumerate}
\item[{[i]}] Can we gain physical insights by identifying the distinguishing
features that characterize the geometry along evolution of pure and mixed
quantum states?

\item[{[ii]}] Expressing the concept of complexity in terms of volumes of
explored regions on curved manifolds, do geodesic paths on manifolds of mixed
quantum states exhibit a higher degree of complexity compared to the
complexity of geodesic paths emerging from the geometry along the evolution of
pure quantum states?

\item[{[iii]}] Does the choice of the metric on the space of mixed quantum
states have crucial observable physical effects on the complexity of the
underlying geodesic paths?
\end{enumerate}

The layout of the rest of the paper is as follows. In Section II, we introduce
our proposed measure of complexity of geodesic paths on curved manifolds. In
Section III, we introduce the geodesic paths on manifolds of pure and mixed
states emerging from the Fubini-Study and the Sj\"{o}qvist metrics,
respectively. In Section IV, we study the complexity of the geodesic paths
expressed in terms of temporal averages of volume regions explored by the
physical systems during the quantum evolutions. In Section V, we include
several physics considerations, including comments on the concepts of metric,
path length, and curvature employed in our analysis. These comments also help
emphasizing the physical significance of our proposed complexity measure. In
Section VI, we present our final remarks. Finally, several technical details,
including a comparative analysis between the Sj\"{o}qvist and the Bures
metrics for mixed quantum states, appear in Appendix A, B, C, D, E, and F.

\section{Information Geometric Complexity}

In this section, we present the notion of information geometric complexity
(IGC) along with the concept of information geometric entropy (IGE). These
quantities will help quantifying how complex are the evolutions of pure and
mixed states. Before introducing formal details, let us emphasize at the
outset that the IGC is essentially the exponential of the IGE. The latter, in
turn, is the logarithm of the volume of the parametric region explored by the
system during its evolution from an initial to a final configuration on the
underlying manifold. The IGE is an indicator of complexity that was initially
proposed in Ref. \cite{cafaro07} in the framework of the Information Geometric
Approach to Chaos (IGAC) \cite{cafarothesis}. For clarity, we mention in this
paper only the necessary information on the IGAC. However, we recommend the
interested reader to consider the compact discussions on the IGAC in Refs.
\cite{ali18,ali21}.

In what follows, we begin by presenting the IGE in its original classical
setting characterized by probability density functions. Obviously, when
transitioning from classical to quantum settings, parametrized families of
probability distributions are replaced by families of parametrized density operators.

Assume that $N$-real valued variables $\left(  \xi^{1}\text{,..., }\xi
^{N}\right)  $ parametrize the points $\left\{  p\left(  x\text{; }\xi\right)
\right\}  $ of an $N$-dimensional curved statistical manifold $\mathcal{M}%
_{s}$,%
\begin{equation}
\mathcal{M}_{s}\overset{\text{def}}{=}\left\{  p\left(  x\text{; }\xi\right)
:\xi\overset{\text{def}}{=}\left(  \xi^{1}\text{,..., }\xi^{N}\right)
\in\mathcal{D}_{\boldsymbol{\xi}}^{\mathrm{tot}}\right\}  \text{.}
\label{unooo}%
\end{equation}
In addition, assume that the microvariables $x$ specifying the probability
distributions $\left\{  p\left(  x\text{; }\xi\right)  \right\}  $ are
elements of the (continuous) microspace $\mathcal{X}$ while the macrovariables
$\xi$ belong to the parameter space $\mathcal{D}_{\boldsymbol{\xi}%
}^{\mathrm{tot}}$ defined as,
\begin{equation}
\mathcal{D}_{\boldsymbol{\xi}}^{\mathrm{tot}}\overset{\text{def}}{=}\left(
\mathcal{I}_{\xi^{1}}\otimes\mathcal{I}_{\xi^{2}}\text{...}\otimes
\mathcal{I}_{\xi^{N}}\right)  \subseteq\mathbb{R}^{N}\text{.} \label{dtot}%
\end{equation}
Note that $\mathcal{I}_{\xi^{j}}$ in Eq. (\ref{dtot}) is a subset of
$\mathbb{R}^{N}$ and characterizes the range of acceptable values for the
statistical macrovariables $\xi^{k}$ with $1\leq k\leq N$. Within the IGAC
framework, it is argued that the IGE is a good measure of temporal complexity
of geodesic paths on $\mathcal{M}_{s}$. The IGE\ is given by,
\begin{equation}
\mathcal{S}_{\mathcal{M}_{s}}\left(  \tau\right)  \overset{\text{def}}{=}%
\log\widetilde{vol}\left[  \mathcal{D}_{\boldsymbol{\xi}}\left(  \tau\right)
\right]  \text{,} \label{IGE}%
\end{equation}
with the average dynamical statistical volume\textbf{\ }$\widetilde{vol}%
\left[  \mathcal{D}_{\boldsymbol{\xi}}\left(  \tau\right)  \right]  $ being
defined as,
\begin{equation}
\widetilde{vol}\left[  \mathcal{D}_{\boldsymbol{\xi}}\left(  \tau\right)
\right]  \overset{\text{def}}{=}\frac{1}{\tau}\int_{0}^{\tau}vol\left[
\mathcal{D}_{\boldsymbol{\xi}}\left(  \tau^{\prime}\right)  \right]
d\tau^{\prime}\text{.} \label{rhs}%
\end{equation}
Note that $\mathcal{D}_{\boldsymbol{\xi}}\left(  \tau^{\prime}\right)  $ in
Eq. (\ref{rhs}) is an $N$-dimensional subspace of $\mathcal{D}%
_{\boldsymbol{\xi}}^{\mathrm{tot}}\subseteq\mathbb{R}^{N}$ whose elements
$\left\{  \xi\right\}  $ with $\xi\overset{\text{def}}{=}\left(  \xi
^{1}\text{,..., }\xi^{N}\right)  $ satisfy $\xi^{k}\left(  \tau_{0}\right)
\leq\xi^{k}\leq\xi^{j}\left(  \tau_{0}+\tau^{\prime}\right)  $ with $\tau_{0}$
denoting the initial value taken by the affine parameter $\tau^{\prime}$ that
characterizes the geodesics on $\mathcal{M}_{s}$ as will be described in more
detail shortly. In Eq. (\ref{rhs}), the temporal average operation is denoted
with the tilde symbol. We also emphasize that\textbf{ }two sequential
integration procedures define $\widetilde{vol}\left[  \mathcal{D}%
_{\boldsymbol{\xi}}\left(  \tau\right)  \right]  $\textbf{ }in Eq.
(\ref{rhs}). The first integration is defined on the explored parameter space
$\mathcal{D}_{\boldsymbol{\xi}}\left(  \tau^{\prime}\right)  $ and leads to
$vol\left[  \mathcal{D}_{\boldsymbol{\xi}}\left(  \tau^{\prime}\right)
\right]  $. Then, the second integration describes a temporal averaging
procedure, is performed over the duration $\tau$ of the evolution on
$\mathcal{M}_{s}$, and finally yields $\widetilde{vol}\left[  \mathcal{D}%
_{\boldsymbol{\xi}}\left(  \tau\right)  \right]  $. The volume\textbf{\ }%
$vol\left[  \mathcal{D}_{\boldsymbol{\xi}}\left(  \tau^{\prime}\right)
\right]  $\textbf{\ }in the RHS of Eq. (\ref{rhs}) is the volume of an
extended region on $\mathcal{M}_{s}$ and is given by,%
\begin{equation}
vol\left[  \mathcal{D}_{\boldsymbol{\xi}}\left(  \tau^{\prime}\right)
\right]  \overset{\text{def}}{=}\int_{\mathcal{D}_{\boldsymbol{\xi}}\left(
\tau^{\prime}\right)  }\rho\left(  \xi^{1}\text{,..., }\xi^{N}\right)
d^{N}\xi\text{.} \label{v}%
\end{equation}
Since we are limiting our present discussion to the IGE in the context of a
statistical manifold $\mathcal{M}_{s}$ of classical probability distributions,
$\rho\left(  \xi^{1}\text{,..., }\xi^{N}\right)  $ in\ Eq. (\ref{v}) is the
so-called Fisher density and equals the square root of the determinant
$g\left(  \xi\right)  $ of the Fisher-Rao information metric tensor $g_{\mu
\nu}^{\mathrm{FR}}\left(  \xi\right)  $, $g^{\mathrm{FR}}\left(  \xi\right)
\overset{\text{def}}{=}\det\left[  g_{\mu\nu}^{\mathrm{FR}}\left(  \xi\right)
\right]  $. Therefore, $\rho\left(  \xi^{1}\text{,..., }\xi^{N}\right)
\overset{\text{def}}{=}\sqrt{g^{\mathrm{FR}}\left(  \xi\right)  }$. Recall
that in the continuous microspace setting, $g_{\mu\nu}^{\mathrm{FR}}\left(
\xi\right)  $ is defined as%
\begin{equation}
g_{\mu\nu}^{\mathrm{FR}}\left(  \xi\right)  \overset{\text{def}}{=}\int
p\left(  x|\xi\right)  \partial_{\mu}\log p\left(  x|\xi\right)  \partial
_{\nu}\log p\left(  x|\xi\right)  dx\text{,} \label{FRmetric}%
\end{equation}
with $\partial_{\mu}\overset{\text{def}}{=}\partial/\partial\xi^{\mu}$. Note
that $vol\left[  \mathcal{D}_{\boldsymbol{\theta}}\left(  \tau^{\prime
}\right)  \right]  $ in Eq. (\ref{v}) assumes a more simple expression for
manifolds equipped with metric tensors specified by factorizable
determinants,
\begin{equation}
g\left(  \xi\right)  =g\left(  \xi^{1}\text{,..., }\xi^{N}\right)
={\prod\limits_{k=1}^{N}}g_{k}\left(  \xi^{k}\right)  \text{.}%
\end{equation}
In such a scenario, the IGE in Eq. (\ref{IGE}) reduces to
\begin{equation}
\mathcal{S}_{\mathcal{M}_{s}}\left(  \tau\right)  =\log\left\{  \frac{1}{\tau
}\int_{0}^{\tau}\left[  {\prod\limits_{k=1}^{N}}\left(  \int_{\tau_{0}}%
^{\tau_{0}+\tau^{\prime}}\sqrt{g_{k}\left[  \xi^{k}\left(  \eta\right)
\right]  }\frac{d\xi^{k}}{d\eta}d\eta\right)  \right]  d\tau^{\prime}\right\}
\text{.} \label{IGEmod}%
\end{equation}
We remark the $g\left(  \theta\right)  $ is not factorizable when the
microvariables $\left\{  x\right\}  $ are correlated. Therefore, in this case,
one is forced to use the general definition of the IGE. We refer to Ref.
\cite{ali10} for a study on the effects of microscopic correlations on the IGE
of Gaussian statistical models.

In the IGAC theoretical setting, the leading asymptotic behavior of
$\mathcal{S}_{\mathcal{M}_{s}}\left(  \tau\right)  $ in Eq. (\ref{IGEmod})
characterizes the complexity of the statistical models being analyzed. To this
end, we consider the leading asymptotic term in the equation for the IGE,
\begin{equation}
\mathcal{S}_{\mathcal{M}_{s}}^{\text{\textrm{asymptotic}}}\left(  \tau\right)
\sim\lim_{\tau\rightarrow\infty}\left[  \mathcal{S}_{\mathcal{M}_{s}}\left(
\tau\right)  \right]  \text{.} \label{LONG}%
\end{equation}
Observe that $\mathcal{D}_{\boldsymbol{\xi}}\left(  \tau^{\prime}\right)  $
specifies the domain of integration that appears in the expression of
$vol\left[  \mathcal{D}_{\boldsymbol{\xi}}\left(  \tau^{\prime}\right)
\right]  $ in Eq. (\ref{v}), and is defined as%
\begin{equation}
\mathcal{D}_{\boldsymbol{\xi}}\left(  \tau^{\prime}\right)
\overset{\text{def}}{=}\left\{  \xi:\xi^{k}\left(  \tau_{0}\right)  \leq
\xi^{k}\leq\xi^{k}\left(  \tau_{0}+\tau^{\prime}\right)  \right\}  \text{,}
\label{L1}%
\end{equation}
where $\tau_{0}\leq\eta\leq\tau_{0}+\tau^{\prime}$ and $\tau_{0}$ is the
initial value of the affine parameter $\eta$. In Eq. (\ref{L1}), $\xi^{k}%
=\xi^{k}\left(  \eta\right)  $ satisfy the geodesic equations
\begin{equation}
\frac{d^{2}\xi^{k}}{d\eta^{2}}+\Gamma_{ij}^{k}\frac{d\xi^{i}}{d\eta}\frac
{d\xi^{j}}{d\eta}=0\text{,} \label{ge}%
\end{equation}
with $\Gamma_{ik}^{j}$ in\ Eq. (\ref{ge}) being the usual Christoffel
connection coefficients,%
\begin{equation}
\Gamma_{ij}^{k}\overset{\text{def}}{=}\frac{1}{2}g^{kl}\left(  \partial
_{i}g_{lj}+\partial_{j}g_{il}-\partial_{l}g_{ij}\right)  \text{.}%
\end{equation}
Note that the elements of $\mathcal{D}_{\boldsymbol{\xi}}\left(  \tau^{\prime
}\right)  $ in Eq. (\ref{L1}), an $N$-dimensional subspace of $\mathcal{D}%
_{\boldsymbol{\xi}}^{\mathrm{tot}}$, are $N$-dimensional macrovariables
$\left\{  \xi\right\}  $ with components $\xi^{j}$ bounded by fixed
integration limits $\xi^{j}\left(  \tau_{0}\right)  $ and $\xi^{j}\left(
\tau_{0}+\tau^{\prime}\right)  $. The temporal functional form of such limits
can be determined by integrating the $N$-coupled nonlinear second order ODEs
in Eq. (\ref{ge}). Having introduced the IGE, we term information geometric
complexity (IGC) the quantity $\mathcal{C}_{\mathcal{M}_{s}}\left(
\tau\right)  $ given by%
\begin{equation}
\mathcal{C}_{\mathcal{M}_{s}}\left(  \tau\right)  \overset{\text{def}%
}{=}\widetilde{vol}\left[  \mathcal{D}_{\boldsymbol{\xi}}\left(  \tau\right)
\right]  =e^{\mathcal{S}_{\mathcal{M}_{s}}\left(  \tau\right)  }\text{.}
\label{IGC}%
\end{equation}
As mentioned earlier, we shall focus on the asymptotic temporal behavior of
the IGC as specified by $\mathcal{C}_{\mathcal{M}_{s}}%
^{\text{\textrm{asymptotic}}}\left(  \tau\right)  \overset{\tau\rightarrow
\infty}{\sim}e^{\mathcal{S}_{\mathcal{M}_{s}}\left(  \tau\right)  }$. \ 

The IGC $\mathcal{C}_{\mathcal{M}_{s}}\left(  \tau\right)  $ can be
interpreted by explaining the meaning of the IGE $\mathcal{S}_{\mathcal{M}%
_{s}}\left(  \tau\right)  $ in Eq. (\ref{IGE}). The IGE is an affine temporal
average of the\textbf{\ }$N$\textbf{-}fold integral of the Fisher density over
geodesics regarded as maximum probability trajectories and, in addition,
measures the number of the explored macrostates in $\mathcal{M}_{s}$. In
particular, the IGE at a given instant is the logarithm of the volume of the
effective parameter space navigated by the system at that specific instant.
The temporal averaging procedure in Eq. (\ref{rhs}) is introduced to average
out the conceivably very complicated fine details of the probabilistic
dynamical description of the system on $\mathcal{M}_{s}$. Furthermore, the
long-time limit in\ Eq. (\ref{LONG}) is used to properly specify the selected
dynamical indicators of complexity by neglecting the transient effects which
enter the calculation of the expected value of the volume of the effective
parameter space. In summary, the IGE provides an asymptotic coarse-grained
inferential characterization of the complex dynamics of a system in the
presence of partial knowledge. For further details on the IGC and IGE, we
refer to Refs. \cite{cafaro18,cafaro17,cafaro10}.

Our discussion has followed the original IGAC setting where we assumed to deal
with an underlying continuous microspace yielding a macrospace \ equipped with
a classical Fisher-Rao information metric in its integral form. However,
shifting to a discrete microspace leading to a macrospace with a Fisher-Rao
information metric expressed in terms of a summation is straightforward,%
\begin{equation}
g_{\mu\nu}^{\mathrm{FR}}\left(  \xi\right)  =\sum_{k=1}^{N}\frac{1}%
{p_{k}\left(  \xi\right)  }\frac{\partial p_{k}\left(  \xi\right)  }%
{\partial\xi^{\mu}}\frac{\partial p_{k}\left(  \xi\right)  }{\partial\xi^{\nu
}}\text{.} \label{gg1}%
\end{equation}
From Eq. (\ref{gg1}), the Fisher-Rao infinitesimal line element
$ds_{\mathrm{FR}}^{2}$ becomes%
\begin{equation}
ds_{\mathrm{FR}}^{2}=g_{\mu\nu}^{\mathrm{FR}}\left(  \xi\right)  d\xi^{\mu
}d\xi^{\nu}=\sum_{k=1}^{N}\frac{dp_{k}^{2}}{p_{k}}\text{,}%
\end{equation}
where $dp_{k}\overset{\text{def}}{=}\left(  \partial_{\mu}p_{k}\right)
d\xi^{\mu}$. Moreover, the parameters $\left\{  \xi^{k}\right\}  _{1\leq k\leq
N}$ were originally viewed in the IGAC\ context as statistical macrovariables
emerging, for instance, as suitable expectation values of the microvariables
of the physical system in the presence of partial knowledge. However, due to
the fact that in principle the IGE can be fully constructed from a geometric
standpoint once the infinitesimal line element $ds^{2}$ is known, its
extension to quantum manifolds of density matrices $\left\{  \rho_{\xi}\left(
x\right)  \right\}  $ specified by a set of parameters $\left\{  \xi\right\}
$ with $\xi\in\mathcal{D}_{\boldsymbol{\xi}}^{\mathrm{tot}}\subseteq
\mathbb{R}^{N}$, including experimentally controllable parameters such as
temperature and magnetic field intensity, is simple as well. For clarity, note
that $\rho_{\xi}\left(  x\right)  \in\mathcal{M}_{s}^{\left(
\text{\textrm{quantum}}\right)  }$ replaces $p_{\xi}\left(  x\right)
\overset{\text{def}}{=}p\left(  x\text{; }\xi\right)  \in\mathcal{M}%
_{s}^{\left(  \text{\textrm{classical}}\right)  }$ with $\mathcal{M}%
_{s}^{\left(  \text{\textrm{classical}}\right)  }$ equal to $\mathcal{M}_{s}$
in\ Eq. (\ref{unooo}).

Clearly, to provide estimates of the IGE and of the IGC in Eqs. (\ref{IGE})
and (\ref{IGC}), respectively, we need to first find the geodesic paths on the
manifolds. Therefore, in the next section, we present the geodesic paths on
the manifolds of pure and mixed states equipped with the Fubini-Study and
Sj\"{o}qvist metrics, respectively.

\section{Geodesic Paths}

We introduce here the geodesic paths on manifolds of pure and mixed states
equipped with the Fubini-Study and the Sj\"{o}qvist metrics, respectively.

\subsection{Geodesic paths on the Bloch sphere: The Fubini-Study metric}

We begin by discussing geodesic paths on manifolds of pure states equipped
with the Fubini-Study metric. In quantum mechanics, it is known that the only
Riemannian metric on the set of rays, up to a constant factor, which is
invariant under all unitary transformations is the angle in Hilbert space
(also known as, the Wootters angle),%
\begin{equation}
\theta_{\mathrm{Wootters}}\left(  \left\vert \psi_{i}\right\rangle \text{,
}\left\vert \psi_{f}\right\rangle \right)  \overset{\text{def}}{=}%
\arccos\left[  \left\vert \left\langle \psi_{i}|\psi_{f}\right\rangle
\right\vert \right]  \text{,} \label{angle1}%
\end{equation}
with $\left\vert \psi_{i}\right\rangle $ and $\left\vert \psi_{f}\right\rangle
$ being two pure states. It is also known that a concept of statistical
distance can be defined between different preparations of the same quantum
system, or to put it another way, between different rays in the same Hilbert
space \cite{wootters81}. This notion of statistical distance is specified
completely by the size of statistical fluctuations taking place in
measurements prepared to discriminate one state from another. A major finding
obtained by Wootters in Ref. \cite{wootters81} was showing that such
statistical distance coincides with the usual distance (i.e., angle) between
rays. The infinitesimal line element that corresponds to the Hilbert space
angle is the so-called Fubini-Study metric $g_{\mu\nu}^{\mathrm{FS}}\left(
\xi\right)  $, the natural metric on the manifold of Hilbert space rays. The
physical interpretation of this metric in terms of statistical fluctuations in
the outcomes of intrinsically probabilistic quantum measurements that aim at
distinguishing one pure state from another is a major result obtained in Ref.
\cite{wootters81}. Before introducing the Fubini-Study metric $g_{\mu\nu
}^{\mathrm{FS}}\left(  \xi\right)  $ in an explicit manner, we remark that the
extension of Wootters' reasoning to the problem of distinguishing mixed
quantum states was carried out by Braunstein and Caves in Ref.
\cite{braunstein94}. In the case of mixed states, the Bures angle
$\theta_{\mathrm{Bures}}$ and the Bures metric $g_{\mu\nu}^{\mathrm{Bures}%
}\left(  \xi\right)  $ replace the Hilbert space angle $\theta
_{\mathrm{Wootters}}$ and the Fubini-Study metric $g_{\mu\nu}^{\mathrm{FS}}$,
respectively. The Bures angle represents the length of a geodesic joining two
density operators $\rho_{i}$ and $\rho_{f}$ and is given by,%
\begin{equation}
\theta_{\mathrm{Bures}}\left(  \rho_{i}\text{, }\rho_{f}\right)
\overset{\text{def}}{=}\arccos\left[  \mathrm{F}_{B}\left(  \rho_{i}\text{,
}\rho_{f}\right)  \right]  \text{.} \label{angle2}%
\end{equation}
In Eq. (\ref{angle2}), $\mathrm{F}_{B}\left(  \rho_{i}\text{, }\rho
_{f}\right)  $ is the Bures fidelity defined as%
\begin{equation}
\mathrm{F}_{B}\left(  \rho_{i}\text{, }\rho_{f}\right)  \overset{\text{def}%
}{=}\left[  \mathrm{Tr}\left(  \sqrt{\rho_{i}^{1/2}\rho_{f}\rho_{i}^{1/2}%
}\right)  \right]  ^{2}\text{.} \label{buref}%
\end{equation}
For clarity, we point out that using Eqs. (\ref{angle2}) and (\ref{buref}),
$\theta_{\mathrm{Bures}}\left(  \rho_{i}\text{, }\rho_{f}\right)  $ equals
$\arccos\left(  \sqrt{\left\langle \psi_{i}|\rho_{f}|\psi_{i}\right\rangle
}\right)  $ when $\rho_{i}\overset{\text{def}}{=}\left\vert \psi
_{i}\right\rangle \left\langle \psi_{i}\right\vert $. Furthermore, when both
$\rho_{i}$ and $\rho_{f}$ are pure states, $\theta_{\mathrm{Bures}}\left(
\rho_{i}\text{, }\rho_{f}\right)  $ reduces to $\theta_{\mathrm{Wootters}%
}\left(  \left\vert \psi_{i}\right\rangle \text{, }\left\vert \psi
_{f}\right\rangle \right)  $ in Eq. (\ref{angle1}). Finally, for completeness,
we remark here that the Bures distance $d_{\mathrm{Bures}}\left(  \rho
_{i}\text{, }\rho_{f}\right)  $ is different from the Bures angle in Eq.
(\ref{angle2}) and is formally defined as%
\begin{equation}
d_{\mathrm{Bures}}\left(  \rho_{i}\text{, }\rho_{f}\right)
\overset{\text{def}}{=}\sqrt{2\left[  1-\sqrt{\mathrm{F}_{B}\left(  \rho
_{i}\text{, }\rho_{f}\right)  }\right]  }\text{.}%
\end{equation}
Returning to the formal introduction of $g_{\mu\nu}^{\mathrm{FS}}\left(
\xi\right)  $, consider two neighboring single qubit pure states $\left\vert
\psi\right\rangle $ and $\left\vert \bar{\psi}\right\rangle $ defined as,%
\begin{equation}
\left\vert \psi\right\rangle \overset{\text{def}}{=}\sum_{k=0}^{1}\sqrt{p_{k}%
}e^{i\phi_{k}}\left\vert e_{k}\right\rangle \text{, and }\left\vert \bar{\psi
}\right\rangle \overset{\text{def}}{=}\sum_{k=0}^{1}\sqrt{p_{k}+dp_{k}%
}e^{i\left(  \phi_{k}+d\phi_{k}\right)  }\left\vert e_{k}\right\rangle
\text{,} \label{g1}%
\end{equation}
respectively, with $\left\{  \left\vert e_{k}\right\rangle \right\}  $ being
an orthonormal basis of the Hilbert space of single qubit state vectors. The
infinitesimal line element between $\left\vert \psi\right\rangle $ and
$\left\vert \bar{\psi}\right\rangle $ in Eq. (\ref{g1}) is given by the
Fubini-Study metric $ds_{\mathrm{FS}}^{2}$ \cite{braunstein94},%
\begin{equation}
ds_{\mathrm{FS}}^{2}\overset{\text{def}}{=}1-\left\vert \left\langle \bar
{\psi}|\psi\right\rangle \right\vert ^{2}=\frac{1}{4}\sum_{k=0}^{1}%
\frac{dp_{k}^{2}}{p_{k}}+\left[  \sum_{k=0}^{1}p_{k}d\phi_{k}^{2}-\left(
\sum_{k=0}^{1}p_{k}d\phi_{k}\right)  ^{2}\right]  \text{.} \label{g2}%
\end{equation}
Using the Bloch sphere parametrization of single qubit states,
\begin{equation}
\left\vert \psi\right\rangle =\left\vert \psi\left(  \theta\text{, }%
\varphi\right)  \right\rangle \overset{\text{def}}{=}\cos\left(  \frac{\theta
}{2}\right)  \left\vert 0\right\rangle +e^{i\varphi}\sin\left(  \frac{\theta
}{2}\right)  \left\vert 1\right\rangle \text{,} \label{g3}%
\end{equation}
where $0\leq\theta\leq\pi$ and $0\leq\varphi<2\pi$, we get by comparing Eqs.
(\ref{g1}) and (\ref{g3}) that%
\begin{equation}
p_{0}\left(  \theta\text{, }\varphi\right)  =\cos^{2}\left(  \frac{\theta}%
{2}\right)  \text{, }p_{1}\left(  \theta\text{, }\varphi\right)  =\sin
^{2}\left(  \frac{\theta}{2}\right)  \text{, }\phi_{0}\left(  \theta\text{,
}\varphi\right)  =0\text{, and }\phi_{1}\left(  \theta\text{, }\varphi\right)
=\varphi\text{.} \label{g4}%
\end{equation}
Therefore, substituting Eq. (\ref{g4}) into Eq. (\ref{g2}), the Fubini-Study
metric $ds_{\mathrm{FS}}^{2}$ reduces to%
\begin{equation}
ds_{\mathrm{FS}}^{2}=g_{\mu\nu}^{\mathrm{FS}}\left(  \xi\right)  d\xi^{\mu
}d\xi^{\nu}=\frac{1}{4}\left[  d\theta^{2}+\sin^{2}\left(  \theta\right)
d\varphi^{2}\right]  \text{.} \label{g4b}%
\end{equation}
In Eq. (\ref{g4b}), $g_{\mu\nu}^{\mathrm{FS}}\left(  \xi\right)  $ is the
Fubini-Study metric tensor, $1\leq\mu$, $\nu\leq2$, and $\xi=\left(  \xi
^{1}\text{, }\xi^{2}\right)  \overset{\text{def}}{=}\left(  \theta\text{,
}\varphi\right)  $. For completeness, we point out that the Fubini-Study
distance between two antipodal (i.e., orthogonal) states on the Bloch sphere
is $\pi/2$. Instead, the geodesic distance between two antipodal states is
$\pi$. Indeed, $ds_{\mathrm{FS}}^{2}=\left(  1/4\right)  ds_{\mathrm{BSM}}%
^{2}$ where $ds_{\mathrm{BSM}}^{2}$ denotes the Bloch sphere metric (BSM)
defined as \cite{uzdin12},%
\begin{equation}
ds_{\mathrm{BSM}}^{2}\overset{\text{def}}{=}d\hat{n}\cdot d\hat{n}\text{.}
\label{g5}%
\end{equation}
In Eq. (\ref{g5}), $\hat{n}$ is the unit vector in $%
%TCIMACRO{\U{211d} }%
%BeginExpansion
\mathbb{R}
%EndExpansion
^{3}$ given by%
\begin{equation}
\hat{n}\overset{\text{def}}{=}\frac{\left\langle \psi\left(  \theta\text{,
}\varphi\right)  |\vec{\sigma}|\psi\left(  \theta\text{, }\varphi\right)
\right\rangle }{\left\langle \psi\left(  \theta\text{, }\varphi\right)
|\psi\left(  \theta\text{, }\varphi\right)  \right\rangle }=\left(
\sin\left(  \theta\right)  \cos\left(  \varphi\right)  \text{, }\sin\left(
\theta\right)  \sin\left(  \varphi\right)  \text{, }\cos\left(  \theta\right)
\right)  \text{,} \label{g5b}%
\end{equation}
with $\vec{\sigma}\overset{\text{def}}{=}\left(  \sigma_{x}\text{, }\sigma
_{y}\text{, }\sigma_{z}\right)  $ being the Pauli vector operator and
$\left\vert \psi\left(  \theta\text{, }\varphi\right)  \right\rangle $ given
in Eq. (\ref{g3}). From Eq. (\ref{g4b}), the only nonvanishing Christoffel
connection coefficients are%
\begin{equation}
\Gamma_{22}^{1}=-\sin\left(  \theta\right)  \cos\left(  \theta\right)  \text{,
and }\Gamma_{12}^{2}=\Gamma_{21}^{2}=\frac{\cos\left(  \theta\right)  }%
{\sin\left(  \theta\right)  }\text{.}%
\end{equation}
Therefore, geodesic paths satisfy the geodesic equations in Eq. (\ref{ge})
being specified by the following system of two coupled second order nonlinear
ODEs,%
\begin{equation}
\ddot{\theta}-\sin\left(  \theta\right)  \cos\left(  \theta\right)
\dot{\varphi}^{2}=0\text{, and }\ddot{\varphi}+2\frac{\cos\left(
\theta\right)  }{\sin\left(  \theta\right)  }\dot{\theta}\dot{\varphi
}=0\text{,} \label{GE1}%
\end{equation}
where $\dot{\theta}\overset{\text{def}}{=}d\theta/d\eta$ with $\eta$ being an
affine parameter. Integration of Eq. (\ref{GE1}) under suitable working
conditions yields geodesic paths given by%
\begin{equation}
\theta\left(  \eta\right)  =\cos^{-1}\left[  \mathrm{a}_{\mathrm{FS}}%
\sin\left(  \eta\right)  \right]  \text{, and }\varphi\left(  \eta\right)
=\varphi_{i}+\tan^{-1}\left[  \mathrm{c}_{\mathrm{FS}}\tan\left(  \eta\right)
\right]  \text{,} \label{geo25}%
\end{equation}
where $\mathrm{a}_{\mathrm{FS}}^{2}\overset{\text{def}}{=}1-\mathrm{c}%
_{\mathrm{FS}}^{2}$ and $\mathrm{c}_{\mathrm{FS}}=\mathrm{c}_{\mathrm{FS}%
}\left(  \theta_{i}\text{, }\dot{\varphi}_{i}\right)  \overset{\text{def}%
}{=}\dot{\varphi}_{i}\sin^{2}\left(  \theta_{i}\right)  =\mathrm{const}$. Note
that both $\theta\left(  \eta\right)  $ and $\varphi\left(  \eta\right)  $ in
Eq. (\ref{geo25}) are bounded functions for any $\eta\geq0$. We remark that
the speed of evolution along these paths is constant and equals
$v_{\mathrm{FS}}\overset{\text{def}}{=}(1/2)\left[  \dot{\theta}^{2}+\sin
^{2}\left(  \theta\right)  \dot{\varphi}^{2}\right]  ^{1/2}$. For a detailed
derivation of the relations in Eq. (\ref{geo25}) along with their extension to
arbitrary working conditions, we refer to Appendix A. Having found the
geodesic paths in Eq. (\ref{geo25}), we focus now on geodesics on manifolds of
mixed quantum states equipped with the Sj\"{o}qvist metric.

\subsection{Geodesic paths in the Bloch ball: The Sj\"{o}qvist metric}

W begin by mentioning the motivation underlying the introduction of the
Sj\"{o}qvist metric from a practical standpoint in science. From a physical
standpoint, the Sj\"{o}qvist metric can be related to measurable quantities in
suitably prepared interferometric measurements. For this reason, it is
sometimes called \textquotedblleft interferometric\textquotedblright\ metric.
The metric can be regarded as the infinitesimal distance $\delta s^{2}\left(
\rho\text{, }\rho+\delta\rho\right)  \approx g\left(  \dot{\rho}\text{, }%
\dot{\rho}\right)  \delta t^{2}$ between two neighboring mixed states $\rho$
and $\rho+\delta\rho$ with $\delta\rho=\dot{\rho}\delta t$. The mixed state
$\rho$ encodes the internal degree of freedom of a particle entering a
Mach-Zehnder interferometer with two beam splitters. The mixed state
$\rho^{\prime}\overset{\text{def}}{=}\rho+\delta\rho$ equals $U\rho
U^{\dagger}$ with $U$ being a unitary applied to the particle for a small but
finite time $\delta t$. From an experimental standpoint, the line element
$\delta s^{2}$ is related to the probability \textrm{P}$_{0}$ of finding the
particle in the $0$-beam (that is, the beam where the unitary transformation
$U$ was applied) after passing the second beam splitter. In particular, up to
the leading nontrivial order in $\delta t$, one finds that \textrm{P}%
$_{0}=1-(1/4)\delta s^{2}$. For more details on a direct experimental access
to the Sj\"{o}qvist line element, we refer to Refs. \cite{erik20,silva21}. In
Ref. \cite{silva21}, the Sj\"{o}qvist metric is generalized by extending its
applicability to degenerate density matrices as well. Interestingly, studying
finite-temperature equilibrium phase transitions, dramatically different
behaviors between the Sj\"{o}qvist and the Bures metrics are noticed in Ref.
\cite{silva21}. Specifically, the Sj\"{o}qvist metric appears to be more
sensitive to the change in parameters than the Bures one. Indeed, unlike what
happens for the Bures metric, the Sj\"{o}qvist metric infers both
zero-temperature and finite-temperature phase transitions. We will return to
this point on the difference between the Sj\"{o}qvist and Bures metrics later
in our paper.

Resuming the formal introduction of the Sj\"{o}qvist metric, consider two
rank-$2$ neighboring nondegenerate density operators $\rho\left(  t\right)  $
and $\rho\left(  t+dt\right)  $ connected via a smooth path $t\mapsto
\rho\left(  t\right)  $ characterizing the evolution of a quantum system. The
nondegeneracy requirement assures that the gauge freedom in the spectral
decomposition of the density operators is represented by the phase of the
eigenvectors. This, in turn, implies there is a one-to-one correspondence
between a rank-$2$ nondegenerate density operator $\rho\left(  t\right)  $ and
the set of two orthogonal rays $\left\{  e^{i\phi_{k}\left(  t\right)
}\left\vert e_{k}\left(  t\right)  \right\rangle :0\leq\phi_{k}\left(
t\right)  <2\pi\right\}  $ that specify the spectral decomposition along the
path $t\mapsto\rho\left(  t\right)  $. Clearly, if some nonzero eigenvalue of
$\rho\left(  t\right)  $ was degenerate, the above mentioned correspondence
would not be valid anylonger. The infinitesimal line element between
$\rho\left(  t\right)  $ and $\rho\left(  t+dt\right)  $ in the working
assumption that $\left\{  \sqrt{p_{k}\left(  t\right)  }e^{i\phi_{k}\left(
t\right)  }\left\vert e_{k}\left(  t\right)  \right\rangle \right\}  _{k=0,1}$
represents the spectral decompositions along the path $t\mapsto\rho\left(
t\right)  $ is given by the Sj\"{o}qvist metric $ds_{\mathrm{Sj\ddot{o}qvist}%
}^{2}$ \cite{erik20},%
\begin{equation}
ds_{\mathrm{Sj\ddot{o}qvist}}^{2}\overset{\text{def}}{=}\min\left[
d^{2}\left(  t\text{, }t+dt\right)  \right]  \text{,}%
\end{equation}
with $d^{2}\left(  t\text{, }t+dt\right)  $ is defined as%
\begin{equation}
d^{2}\left(  t\text{, }t+dt\right)  \overset{\text{def}}{=}\sum_{k=0}%
^{1}\left\Vert \sqrt{p_{k}\left(  t\right)  }e^{i\phi_{k}\left(  t\right)
}\left\vert e_{k}\left(  t\right)  \right\rangle -\sqrt{p_{k}\left(
t+dt\right)  }e^{i\phi_{k}\left(  t+dt\right)  }\left\vert e_{k}\left(
t+dt\right)  \right\rangle \right\Vert ^{2}\text{.}%
\end{equation}
Following the line of reasoning in Ref. \cite{erik20}, $ds_{\mathrm{Sj\ddot
{o}qvist}}^{2}$ can be recast as
\begin{equation}
ds_{\mathrm{Sj\ddot{o}qvist}}^{2}=\frac{1}{4}\sum_{k=0}^{1}\frac{dp_{k}^{2}%
}{p_{k}}+\sum_{k=0}^{1}p_{k}ds_{k}^{2}\text{,} \label{g6}%
\end{equation}
where $dp_{k}=\dot{p}_{k}dt$ and, recalling Ref. \cite{provost80}, $ds_{k}%
^{2}$ in Eq. (\ref{g6}) is the Fubini-Study metric along the pure state
$\left\vert e_{k}\right\rangle $%
\begin{equation}
ds_{k}^{2}\overset{\text{def}}{=}\left\langle \dot{e}_{k}\left\vert \left(
\hat{1}-\left\vert e_{k}\right\rangle \left\langle e_{k}\right\vert \right)
\right\vert \dot{e}_{k}\right\rangle =\left\langle de_{k}|de_{k}\right\rangle
-\left\vert \left\langle e_{k}|de_{k}\right\rangle \right\vert ^{2}\text{,}%
\end{equation}
with $\hat{1}$ being the identity operator on the Hilbert space of single
qubit quantum states. Using the Bloch sphere parametrization of single qubit
mixed states in the Bloch ball, we have%
\begin{equation}
\rho=\frac{\hat{1}+\vec{r}\cdot\vec{\sigma}}{2}=\frac{1}{2}\left(
\begin{array}
[c]{cc}%
1+r\cos\left(  \theta\right)  & r\sin\left(  \theta\right)  e^{-i\varphi}\\
r\sin\left(  \theta\right)  e^{i\varphi} & 1-r\cos\left(  \theta\right)
\end{array}
\right)  \text{,} \label{g7}%
\end{equation}
where $\vec{r}$ is the polarization vector given by $\vec{r}%
\overset{\text{def}}{=}r\hat{n}$ with $\hat{n}$ defined in Eq. (\ref{g5b}).
Note that for mixed quantum states, $0\leq r<1$ and $\det\left(  \rho\right)
=\left(  1/2\right)  \left(  1-\vec{r}^{2}\right)  \geq0$ because of the
positiveness of $\rho$. For pure quantum states, instead, $r=1$ and
$\det\left(  \rho\right)  =0$. From Eq. (\ref{g7}), we observe that the
spectral decomposition of $\rho$ is given by%
\begin{equation}
\rho=\sum_{k=0}^{1}p_{k}\left\vert e_{k}\right\rangle \left\langle
e_{k}\right\vert \text{.}%
\end{equation}
The two distinct eigenvalues $\left\{  p_{k}\right\}  _{k=0,1}$ are given by,%
\begin{equation}
p_{0}=p_{0}\left(  r\text{, }\theta\text{, }\varphi\right)
\overset{\text{def}}{=}\frac{1+r}{2}\text{, and }p_{1}=p_{1}\left(  r\text{,
}\theta\text{, }\varphi\right)  \overset{\text{def}}{=}\frac{1-r}{2}\text{,}
\label{g8}%
\end{equation}
respectively. The orthonormal eigenvectors corresponding to $p_{0}$ and
$p_{1}$ in Eq. (\ref{g8}) are
\begin{equation}
\left\vert e_{0}\right\rangle =\left\vert e_{0}\left(  r\text{, }\theta\text{,
}\varphi\right)  \right\rangle \overset{\text{def}}{=}\frac{1}{\sqrt{2}%
}\left(
\begin{array}
[c]{c}%
e^{-i\varphi}\sqrt{1+\cos\left(  \theta\right)  }\\
\frac{\sin\left(  \theta\right)  }{\sqrt{1+\cos\left(  \theta\right)  }}%
\end{array}
\right)  =\left(
\begin{array}
[c]{c}%
e^{-i\varphi}\cos\left(  \frac{\theta}{2}\right) \\
\sin\left(  \frac{\theta}{2}\right)
\end{array}
\right)  \text{,} \label{g9}%
\end{equation}
and,%
\begin{equation}
\left\vert e_{1}\right\rangle =\left\vert e_{1}\left(  r\text{, }\theta\text{,
}\varphi\right)  \right\rangle \overset{\text{def}}{=}\frac{1}{\sqrt{2}%
}\left(
\begin{array}
[c]{c}%
-e^{-i\varphi}\sqrt{1-\cos\left(  \theta\right)  }\\
\frac{\sin\left(  \theta\right)  }{\sqrt{1-\cos\left(  \theta\right)  }}%
\end{array}
\right)  =\left(
\begin{array}
[c]{c}%
-e^{-i\varphi}\sin\left(  \frac{\theta}{2}\right) \\
\cos\left(  \frac{\theta}{2}\right)
\end{array}
\right)  \text{,} \label{g10}%
\end{equation}
respectively. Finally, using Eqs. (\ref{g8}), (\ref{g9}), and (\ref{g10}),
$ds_{\mathrm{Sj\ddot{o}qvist}}^{2}$ in Eq. (\ref{g6}) becomes%
\begin{equation}
ds_{\mathrm{Sj\ddot{o}qvist}}^{2}=g_{\mu\nu}^{\mathrm{Sj\ddot{o}qvist}}\left(
\xi\right)  d\xi^{\mu}d\xi^{\nu}=\frac{1}{4}\left[  \frac{dr^{2}}{1-r^{2}%
}+d\Omega^{2}\right]  \text{,} \label{g11}%
\end{equation}
with $d\Omega^{2}\overset{\text{def}}{=}d\theta^{2}+\sin^{2}\left(
\theta\right)  d\varphi^{2}$. In Eq. (\ref{g11}), $g_{\mu\nu}^{\mathrm{Sj\ddot
{o}qvist}}\left(  \xi\right)  $ is the Sj\"{o}qvist metric tensor, $1\leq\mu$,
$\nu\leq3$, and $\xi=\left(  \xi^{1}\text{, }\xi^{2}\text{, }\xi^{3}\right)
\overset{\text{def}}{=}\left(  r\text{, }\theta\text{, }\varphi\right)  $.
Note when $r$ is constant and equals one, $ds_{\mathrm{Sj\ddot{o}qvist}}^{2}$
in Eq. (\ref{g11}) reduces to $ds_{\mathrm{FS}}^{2}$ in Eq. (\ref{g4b}). For
completeness, we recall that the Bures metric extends to mixed quantum states
the Fubini-Study metric on pure states \cite{bures69,uhlmann76,hubner92}.
Furthermore, as shown in Ref. \cite{braunstein94}, it is equivalent, up to a
proportionality factor of four, to the quantum Fisher information metric.
Interestingly, we remark that the Bures infinitesimal line element
$ds_{\mathrm{Bures}}^{2}$ between $\rho$ and $\rho+d\rho$ with $\rho$ given in
Eq. (\ref{g7}) is given by%
\begin{equation}
ds_{\mathrm{Bures}}^{2}=g_{\mu\nu}^{\mathrm{Bures}}\left(  \xi\right)
d\xi^{\mu}d\xi^{\nu}=\frac{1}{4}\left[  \frac{dr^{2}}{1-r^{2}}+r^{2}%
d\Omega^{2}\right]  \text{.} \label{g12}%
\end{equation}
For an explicit derivation of Eq. (\ref{g12}), we refer to Appendix B. From
Eqs. (\ref{g11}) and (\ref{g12}), we notice that the angular part of
$ds_{\mathrm{Sj\ddot{o}qvist}}^{2}$ does not exhibit the $r^{2}$-factor which,
instead, appears in $ds_{\mathrm{Bures}}^{2}$. The lack of this factor implies
that the Sj\"{o}qvist metric is singular at the origin of the Bloch ball where
$r=0$ and, unlike the Bures metric, is not defined for degenerate density
operators. Finally, we refer to Ref. \cite{silva21} for a recent extension of
the Sj\"{o}qvist metric for the space of nondegenerate density matrices, to
the degenerate case, i.e., the case in which the eigenspaces have dimension
greater than or equal to one. We will go back to this point on the difference
between the Sj\"{o}qvist and Bures metrics later in our paper (see also
Appendix C and Appendix D).

Returning to the Sj\"{o}qvist metric analysis, we see from Eq. (\ref{g11})
that the only nonvanishing Christoffel connection coefficients are%
\begin{equation}
\Gamma_{11}^{1}=\frac{r}{1-r^{2}}\text{, }\Gamma_{33}^{2}=-\sin\left(
\theta\right)  \cos\left(  \theta\right)  \text{, and }\Gamma_{23}^{3}%
=\Gamma_{32}^{3}=\frac{\cos\left(  \theta\right)  }{\sin\left(  \theta\right)
}\text{.}%
\end{equation}
Therefore, geodesics satisfy the geodesic equations in Eq. (\ref{ge})
described in terms of a system of three coupled second order nonlinear ODEs,%
\begin{equation}
\ddot{r}+\frac{r}{1-r^{2}}\dot{r}^{2}=0\text{, }\ddot{\theta}-\sin\left(
\theta\right)  \cos\left(  \theta\right)  \dot{\varphi}^{2}=0\text{, and
}\ddot{\varphi}+2\frac{\cos\left(  \theta\right)  }{\sin\left(  \theta\right)
}\dot{\theta}\dot{\varphi}=0\text{,} \label{GE2}%
\end{equation}
where $\dot{r}\overset{\text{def}}{=}dr/d\eta$ with $\eta$ being an affine
parameter. Interestingly, observe that although the ODE satisfied by the
radial parameter $r$ in Eq. (\ref{GE2}) is nonlinear, it is not coupled to the
ODEs describing the evolution of the angular parameters $\theta$ and $\varphi
$. Furthermore, the angular motion is identical to the one that emerges when
employing the Fubini-Study metric. Therefore, we refer to Eq. (\ref{geo25})
and to Appendix A for details on the angular motion. Instead, integration of
the radial equation of motion in Eq. (\ref{GE2}) yields,%
\begin{equation}
r_{\mathrm{Sj\ddot{o}qvist}}\left(  \eta\right)  =\sin\left[  \sin^{-1}\left(
r_{i}\right)  +\frac{\dot{r}_{i}}{\sqrt{1-r_{i}^{2}}}\eta\right]  \text{,}
\label{RERIK}%
\end{equation}
where $r_{i}\overset{\text{def}}{=}r\left(  \eta_{i}\right)  $, $\dot{r}%
_{i}\overset{\text{def}}{=}\dot{r}\left(  \eta_{i}\right)  $, and $\eta_{i}$
is set equal to zero. We emphasize that the speed of evolution along geodesic
paths is constant and equals $v_{\mathrm{Sj\ddot{o}qvist}}\overset{\text{def}%
}{=}(1/2)\left[  \left(  1-r^{2}\right)  ^{-1}\dot{r}^{2}+\dot{\theta}%
^{2}+\sin^{2}\left(  \theta\right)  \dot{\varphi}^{2}\right]  ^{1/2}$. For an
explicit derivation of Eq. (\ref{RERIK}) along with a discussion on
alternative geodesic parametrizations like the one used in Ref. \cite{erik20},
we refer to\ Appendix C. Finally, for a discussion on the integration of the
geodesic equations in a Bloch ball equipped with the Bures metric $g_{\mu\nu
}^{\mathrm{Bures}}\left(  \xi\right)  $, we refer to Appendix D. In Appendix
E, instead, we present a summary of curvature properties of the manifold of
pure states equipped with $g_{\mu\nu}^{\mathrm{FS}}\left(  \xi\right)  $ along
with those of a manifold of mixed quantum states endowed with $g_{\mu\nu
}^{\mathrm{Sj\ddot{o}qvist}}\left(  \xi\right)  $ and $g_{\mu\nu
}^{\mathrm{Bures}}\left(  \xi\right)  $. More specifically, for each scenario,
we find the expressions of the tensor metric components, infinitesimal line
elements, Christoffel connection coefficients, Ricci tensor components,
Riemann curvature tensor components, scalar curvatures and, finally, sectional curvatures.

At this point, having found the geodesic paths on curved manifolds equipped
with the Fubini-Study and Sj\"{o}qvist metrics, we are ready to use our
complexity quantifiers in Eqs. (\ref{IGE}) and (\ref{IGC}) to determine how
complex evolutions on pure and mixed states are.

\section{Complexity of quantum evolution}

We study here the complexity of the geodesic paths expressed in terms of
temporal averages of volume regions explored by the physical systems during
the quantum evolutions.

\subsection{Actions, lengths, and accessible volumes}

Before studying the complexity, let us first comment on the relevance of the
concepts of length and action in the geometric formulation of physical
theories. In the Introduction, we mentioned these concepts play a key role in
the understanding of the physics of black holes \cite{leo14,leo15,leo16,
leo16B}. However, lengths and actions also play a very important role in the
geometric formulation of thermodynamics \cite{orlando22}. In this case, these
two quantities are generally termed thermodynamic length and thermodynamic
divergence, respectively. Indeed, including the theory of fluctuations into
the axioms of equilibrium thermodynamics \cite{ruppeiner79}, thermodynamic
systems can be characterized by Riemannian manifolds furnished of a
thermodynamic metric tensor that is identical to the Fisher-Rao information
metric \cite{crooks07}. Within this geometric setting for thermodynamics, the
above mentioned Riemannian structure allows one to introduce a notion of
length for fluctuations about equilibrium states as well as for thermodynamic
processes proceeding via equilibrium states. In analogy to Wootters'
statistical distance between probability distributions as presented in Ref.
\cite{wootters81}, the thermodynamic length of a path connecting two points on
a manifold of thermal states can be viewed as a measure of the maximal number
of statistically distinguishable thermodynamic states along the path
\cite{diosi84}. In particular, the larger the fluctuations, the closer the
points are together. The thermodynamic divergence of a path, instead, is a
measure of the losses in the process quantified by the total entropy produced
along the path. For more details, we refer to Refs.
\cite{ruppeiner79,weinhold75,salamon83}. Having in mind Wootters' approach,
note that the concepts of action and length are formally different when
studying the geometry along the evolution of states. For the sake of
reasoning, assume that the line element of the Riemannian space is given by
$ds^{2}\overset{\text{def}}{=}g_{\mu\nu}\left(  \xi\right)  d\xi^{\mu}%
d\xi^{\nu}$. Then, the action \textrm{A }is given by%
\begin{equation}
\mathrm{A}\overset{\text{def}}{=}\frac{1}{2}m\int_{0}^{\tau}g_{\mu\nu}\left(
\xi\right)  \dot{\xi}^{\mu}\dot{\xi}^{\nu}d\eta\text{,}%
\end{equation}
with $\dot{\xi}\overset{\text{def}}{=}d\xi/d\eta$. The length $\mathcal{L}$ of
a path $\xi^{\mu}\left(  \eta\right)  $ with $0\leq\eta\leq\tau$, instead, is
defined as%
\begin{equation}
\mathcal{L}\overset{\text{def}}{=}\int_{0}^{\tau}\sqrt{g_{\mu\nu}\left(
\xi\right)  \dot{\xi}^{\mu}\dot{\xi}^{\nu}}d\eta\text{.} \label{length}%
\end{equation}
However, for particles of mass $m$ moving along geodesics with constant
velocity, both velocity and energy are conserved. In this case, the path
$\mathcal{L}$ and the action $\mathrm{A}$ are linearly related. Indeed, one
has%
\begin{equation}
\mathrm{A}=\sqrt{\frac{mE}{2}}\mathcal{L}\text{,}%
\end{equation}
where $E\overset{\text{def}}{=}\left(  1/2\right)  mv^{2}$ and $v^{2}%
\overset{\text{def}}{=}g_{\mu\nu}\left(  \xi\right)  \dot{\xi}^{\mu}\dot{\xi
}^{\nu}$ are both constant. Interestingly, we remark that while the length
$\mathcal{L}$ is invariant under reparametrization of the affine parameter
$\eta$, the action $\mathrm{A}$ is not. Before proceeding with the calculation
of lengths in Eq. (\ref{length}) and volumes of explored regions in Eq.
(\ref{rhs}) yielding the complexity of geodesic paths on manifolds of quantum
states, we make a couple of remarks that can help our intuition when
considering the Sj\"{o}qvist and Bures cases with pure calculations. First,
considering Eqs. (\ref{g11}) and (\ref{g12}) while performing a change of
variables defined by $r\overset{\text{def}}{=}\sin\left(  \alpha_{r}\right)  $
with $0\leq\alpha_{r}\leq\pi/2$, we find that $4ds_{\mathrm{Sj\ddot{o}qvist}%
}^{2}=d\alpha_{r}^{2}+d\Omega_{\text{sphere}}^{2}$ and $4ds_{\mathrm{Bures}%
}^{2}=d\alpha_{r}^{2}+\sin^{2}\left(  \alpha_{r}\right)  d\Omega
_{\text{sphere}}^{2}$ with $d\Omega_{\text{sphere}}^{2}\overset{\text{def}%
}{=}d\theta^{2}+\sin^{2}\left(  \theta\right)  d\varphi^{2}$. The structure of
the Sj\"{o}qvist line element recast in this new form is reminiscent of the
structure of a line element in the usual cylindrical coordinates $\left(
\rho\text{, }\varphi\text{, }z\right)  $, $ds_{\text{cylinder}}^{2}%
=dz^{2}+d\Omega_{\text{cylinder}}^{2}$ with $d\Omega_{\text{cylinder}}%
^{2}\overset{\text{def}}{=}d\rho^{2}+\rho^{2}d\varphi^{2}$, once one
identifies the pair $\left(  \alpha_{r}\text{, }d\Omega_{\text{sphere}%
}\right)  $ with the pair $\left(  \rho\text{, }d\Omega_{\text{cylinder}%
}\right)  $. Therefore, one can imagine associating a cylinder with a constant
(varying) radius to the Sj\"{o}qvist (Bures) geometry, respectively. Note that
the varying radius in the Bures case is upper bounded by the constant value
that specifies the radius in the Sj\"{o}qvist geometry. Second, the volumes of
the accessible regions of the manifolds in the Sj\"{o}qvist and Bures
scenarios are given by,%
\begin{equation}
V_{\mathrm{Sj\ddot{o}qvist}}^{\left(  \text{accessible}\right)  }%
\overset{\text{def}}{=}\frac{1}{8}\int_{0}^{1}\int_{0}^{\pi}\int_{0}^{2\pi
}\frac{\sin\left(  \theta\right)  }{\sqrt{1-r^{2}}}drd\theta d\varphi
=\frac{\pi^{2}}{4}\text{,} \label{volo1}%
\end{equation}
and,%
\begin{equation}
V_{\mathrm{Bures}}^{\left(  \text{accessible}\right)  }\overset{\text{def}%
}{=}\frac{1}{8}\int_{0}^{1}\int_{0}^{\pi}\int_{0}^{2\pi}\frac{r^{2}\sin\left(
\theta\right)  }{\sqrt{1-r^{2}}}drd\theta d\varphi=\frac{\pi^{2}}{8}\text{,}
\label{volo2}%
\end{equation}
respectively. Clearly, from Eqs. (\ref{volo1}) and (\ref{volo2}), we note that
$V_{\mathrm{Bures}}^{\left(  \text{accessible}\right)  }\leq
V_{\mathrm{Sj\ddot{o}qvist}}^{\left(  \text{accessible}\right)  }$. This fact,
in turn, is compatible with our intuitive picture proposed in our first
remark. We remark that it is possible that only parts of the accessible
geometric regions are indeed explored during the evolution. In what follows,
we shall finally calculate the lengths and the volumes of the effectively
explored regions of the manifolds in the (pure) Fubini-Study and (mixed)
Sj\"{o}qvist cases.

\subsection{Evolution on the Bloch sphere}

\subsubsection{Length}

In what follows, we focus on calculating the length of geodesics in the unit
Bloch ball that lay int the $xz$-plane specified by the condition $\varphi=0$.
In the Fubini-Study metric case, we have \
\begin{equation}
\mathcal{L}_{\mathrm{FS}}\left(  \eta_{f}\right)  \overset{\text{def}}{=}%
\frac{1}{2}\int_{0}^{\eta_{f}}\frac{d\theta}{d\eta}d\eta=\frac{1}{2}%
\dot{\theta}_{i}\eta_{f}\text{,} \label{lf1}%
\end{equation}
or, alternatively, in terms of the angular variable $\theta$,%
\begin{equation}
\mathcal{L}_{\mathrm{FS}}\left(  \theta_{f}\right)  \overset{\text{def}%
}{=}\frac{\theta_{f}}{2}\text{.} \label{lf2}%
\end{equation}
Note that $\mathcal{L}_{\mathrm{FS}}$ in Eq. (\ref{lf2}) denotes the
Fubini-Study distance ($\theta_{f}/2$) and is half the geodesic distance
($\theta_{f}$) on the Bloch sphere. For completeness, we emphasize that in
obtaining Eq. (\ref{lf1}) we exploited the relation $\ddot{\theta}=0$. This
relation can be obtained from Eq. (\ref{GE1}) once one imposes the constraint
of constant $\varphi$. Moreover, in getting Eq. (\ref{lf2}), we assumed
$\theta_{i}\overset{\text{def}}{=}\theta\left(  \eta_{i}\right)  =0$, with
$\eta_{i}=0$.

\subsubsection{Complexity}

From the Fubini-Study metric $ds_{\mathrm{FS}}^{2}$ in Eq. (\ref{g4b}), we
note that $g_{\mathrm{FS}}\left(  \theta\text{, }\varphi\right)  =\sin
^{2}\left(  \theta\right)  /16$ with $g_{\mathrm{FS}}$ denoting the
determinant of the metric tensor $g_{\mu\nu}^{\mathrm{FS}}$. Therefore, using
Eq. (\ref{geo25}), the instantaneous explored volume region $V_{\mathrm{FS}%
}\left(  \eta\right)  $ as defined in Eq. (\ref{v}) becomes%
\begin{equation}
V_{\mathrm{FS}}\left(  \eta\right)  =\frac{1}{4}\mathrm{a}_{\mathrm{FS}}%
\sin\left(  \eta\right)  \arctan\left[  \mathrm{c}_{\mathrm{FS}}\tan\left(
\eta\right)  \right]  \text{.} \label{iv1}%
\end{equation}
Recall that $\mathrm{a}_{\mathrm{FS}}^{2}\overset{\text{def}}{=}%
1-\mathrm{c}_{\mathrm{FS}}^{2}$ with $\mathrm{c}_{\mathrm{FS}}=\mathrm{c}%
_{\mathrm{FS}}\left(  \theta_{i}\text{, }\dot{\varphi}_{i}\right)
\overset{\text{def}}{=}\dot{\varphi}_{i}\sin^{2}\left(  \theta_{i}\right)
=\mathrm{const}$.\ Then, setting $\theta_{i}=\pi/2$ for illustrative purposes,
we note that the modulus of the volume of the explored region of the manifold
of pure states in Eq. (\ref{iv1}) is upper bounded by $\pi/8$, $\left\vert
V_{\mathrm{FS}}\left(  \eta\right)  \right\vert \leq\pi/8$. Therefore, at a
time $\eta$, less than one eight of the accessible region of the manifold is
actually explored since $V_{\mathrm{FS}}^{\left(  \text{accessible}\right)
}=\pi$. From Eq. (\ref{iv1}), the average explored region as given in Eq.
(\ref{rhs}) represents the IGC in Eq. (\ref{IGC}) and turns into%
\begin{equation}
\mathcal{C}_{\mathrm{FS}}\left(  \tau\right)  =\frac{1}{4}\mathrm{a}%
_{\mathrm{FS}}\frac{I_{V_{\mathrm{FS}}}\left(  \tau\right)  }{\tau}\text{.}
\label{com1}%
\end{equation}
The function $I_{V_{\mathrm{FS}}}\left(  \tau\right)  $ in Eq. (\ref{com1}) is
defined as the integral of $V_{\mathrm{FS}}\left(  \eta\right)  $ with
$0\leq\eta\leq\tau$ and is given by,%
\begin{equation}
I_{V_{\mathrm{FS}}}\left(  \tau\right)  \overset{\text{def}}{=}\frac
{\mathrm{c}_{\mathrm{FS}}}{\sqrt{\mathrm{c}_{\mathrm{FS}}^{2}-1}}%
\arctan\left[  \sqrt{\mathrm{c}_{\mathrm{FS}}^{2}-1}\sin\left(  \tau\right)
\right]  -\cos\left(  \tau\right)  \arctan\left[  \mathrm{c}_{\mathrm{FS}}%
\tan\left(  \tau\right)  \right]  \text{.} \label{funny}%
\end{equation}
The above mentioned integral was performed with the help of the Mathematica
software. Furthermore, we remark that $I_{V_{\mathrm{FS}}}\left(  \tau\right)
$ is a bounded function for any $\tau\geq0$. The asymptotic temporal
expression of the IGC $\mathcal{C}_{\mathrm{FS}}\left(  \tau\right)  $ in Eq.
(\ref{com1}) will be compared with the one that we obtain in the case of mixed
state evolutions with distinguishability metric provided by $g_{\mu\nu
}^{\mathrm{Sj\ddot{o}qvist}}\left(  \xi\right)  $.

\subsection{Evolution in the Bloch ball}

\subsubsection{Length}

In what follows, we focus on calculating the length of geodesics in the unit
Bloch ball that lay int the $xz$-plane specified by the condition $\varphi=0$.
In the Sj\"{o}qvist metric case, recall that $r^{\prime2}\left(
1-r^{2}\right)  ^{-1}=\mathrm{const.}$and $\ddot{\theta}=0$. Therefore, the
length $\mathcal{L}_{\mathrm{Sj\ddot{o}qvist}}\left(  \eta_{f}\right)  $ is
given by%
\begin{equation}
\mathcal{L}_{\mathrm{Sj\ddot{o}qvist}}\left(  \eta_{f}\right)
\overset{\text{def}}{=}\frac{1}{2}\int_{0}^{\eta_{f}}\sqrt{1+\frac{r^{\prime
2}}{1-r^{2}}}\frac{d\theta}{d\eta}d\eta=\frac{1}{2}\sqrt{1+\frac{r_{i}%
^{\prime2}}{1-r_{i}^{2}}}\dot{\theta}_{i}\eta_{f}\text{,} \label{sl1}%
\end{equation}
or, alternatively,%
\begin{equation}
\mathcal{L}_{\mathrm{Sj\ddot{o}qvist}}\left(  \theta_{f}\right)
\overset{\text{def}}{=}\frac{1}{2}\sqrt{\theta_{f}^{2}+\left[  \sin
^{-1}\left(  r_{f}\right)  -\sin^{-1}\left(  r_{i}\right)  \right]  ^{2}%
}\text{.} \label{sl2}%
\end{equation}
Comparing Eqs. (\ref{sl2}) and (\ref{lf2}), we note that $\mathcal{L}%
_{\mathrm{Sj\ddot{o}qvist}}$ reduces to $\mathcal{L}_{\mathrm{FS}}$ when
$r_{f}=r_{i}=1$. For $r_{f}\neq r_{i}$, we generally have%
\begin{equation}
\mathcal{L}_{\mathrm{Sj\ddot{o}qvist}}\left(  \theta_{f}\right)
\geq\mathcal{L}_{\mathrm{FS}}\left(  \theta_{f}\right)  \text{.} \label{ine1}%
\end{equation}
Eq. (\ref{ine1}) implies that the length of a geodesic path connecting two
arbitrary points $P_{i}\overset{\text{def}}{=}\left(  r_{i}\text{, }\theta
_{i}\text{, }0\right)  $ and $P_{f}\overset{\text{def}}{=}\left(  r_{f}\text{,
}\theta_{f}\text{, }0\right)  $ with $r_{i}\neq r_{f}$ in the Bloch ball
laying in the $xz$-plane (i.e., $\varphi=0$) is longer than the length of two
arbitrary points $\tilde{P}_{i}\overset{\text{def}}{=}\left(  \theta
_{i}\text{, }0\right)  $ and $\tilde{P}_{f}\overset{\text{def}}{=}\left(
\theta_{f}\text{, }0\right)  $ laying on the Bloch sphere with $r_{i}=r_{f}=1$
that intercepts the $xz$-plane (i.e., $\varphi=0$). Eq. (\ref{ine1}) hints to
what might happen when comparing the information geometric complexity of the
evolutions of pure and quantum states as we shall see shortly. Complexities
are expressed in terms of volumes. In parametric spaces of dimension higher
than one, relations between length and volumes are not straightforward.
Therefore, we could not easily take the hint as a \textquotedblleft
proof\textquotedblright\ of the higher complexity of the evolution of mixed
quantum states. For this reason, we actually estimate this quantity in an
explicit manner in the following subsection.

\subsubsection{Complexity}

From the Sj\"{o}qvist metric $ds_{\mathrm{Sj\ddot{o}qvist}}^{2}$ in Eq.
(\ref{g11}), we observe that the determinant of the metric tensor $g_{\mu\nu
}^{\mathrm{Sj\ddot{o}qvist}}$ satisfies the relation%
\begin{equation}
g_{\mathrm{Sj\ddot{o}qvist}}\left(  r\text{, }\theta\text{, }\varphi\right)
=\frac{1}{64}\frac{\sin^{2}\left(  \theta\right)  }{1-r^{2}}\text{.}%
\end{equation}
Therefore, making use of Eqs. (\ref{RERIK}) and (\ref{geo25}), the
instantaneous explored volume region $V_{\mathrm{Sj\ddot{o}qvist}}\left(
\eta\right)  $ as given in Eq. (\ref{v}) turns into%
\begin{equation}
V_{\mathrm{Sj\ddot{o}qvist}}\left(  \eta\right)  =\frac{1}{8}\mathrm{a}%
_{\mathrm{FS}}\frac{\dot{r}_{i}}{\sqrt{1-r_{i}^{2}}}\eta\sin\left(
\eta\right)  \arctan\left[  \mathrm{c}_{\mathrm{FS}}\tan\left(  \eta\right)
\right]  \text{.} \label{iv2}%
\end{equation}
Recall that $\mathrm{a}_{\mathrm{FS}}^{2}\overset{\text{def}}{=}%
1-\mathrm{c}_{\mathrm{FS}}^{2}$ with $\mathrm{c}_{\mathrm{FS}}=\mathrm{c}%
_{\mathrm{FS}}\left(  \theta_{i}\text{, }\dot{\varphi}_{i}\right)
\overset{\text{def}}{=}\dot{\varphi}_{i}\sin^{2}\left(  \theta_{i}\right)
=\mathrm{const}$.\ Then, putting $\theta_{i}=\pi/2$ for simplicity, we observe
that the modulus $\left\vert V_{\mathrm{Sj\ddot{o}qvist}}\left(  \eta\right)
\right\vert $ of the volume of the explored region of the manifold of mixed
states in Eq. (\ref{iv2}) is upper bounded by a function that grows linearly
with $\eta$ with proportionality coefficient given by $\dot{r}_{i}%
/\sqrt{1-r_{i}^{2}\text{ }}$ and not by a constant function as in the case of
the evolution of pure quantum states. Moreover, recalling that
$V_{\mathrm{Sj\ddot{o}qvist}}^{\left(  \text{accessible}\right)  }=\pi
^{2}/4\leq\pi=V_{\mathrm{FS}}^{\left(  \text{accessible}\right)  }$, we
clearly expect a more complex behavior for sufficiently large values of $\eta$
in the case of the geometry along evolution of mixed states since%
\begin{equation}
\frac{V_{\mathrm{Sj\ddot{o}qvist}}^{\left(  \text{explored}\right)  }\left(
\eta\right)  }{V_{\mathrm{Sj\ddot{o}qvist}}^{\left(  \text{accessible}\right)
}}\geq\frac{V_{\mathrm{FS}}^{\left(  \text{explored}\right)  }\left(
\eta\right)  }{V_{\mathrm{FS}}^{\left(  \text{accessible}\right)  }}\text{.}
\label{compare}%
\end{equation}
From Eq. (\ref{iv2}), the average explored region as given in Eq. (\ref{rhs})
denotes the IGC in Eq. (\ref{IGC}) and becomes%
\begin{equation}
\mathcal{C}_{\mathrm{Sj\ddot{o}qvist}}\left(  \tau\right)  =\frac{1}{\tau}%
\int_{0}^{\tau}V_{\mathrm{Sj\ddot{o}qvist}}\left(  \eta\right)  d\eta\text{,}%
\end{equation}
that is,%
\begin{equation}
\mathcal{C}_{\mathrm{Sj\ddot{o}qvist}}\left(  \tau\right)  =\frac{1}%
{8}\mathrm{a}_{\mathrm{FS}}\frac{\dot{r}_{i}}{\sqrt{1-r_{i}^{2}}}\frac{1}%
{\tau}\int_{0}^{\tau}\eta\sin\left(  \eta\right)  \arctan\left[
\mathrm{c}_{\mathrm{FS}}\tan\left(  \eta\right)  \right]  d\eta\text{.}
\label{cifro}%
\end{equation}
Denoting $f\left(  \eta\right)  \overset{\text{def}}{=}\eta$ and $\dot
{g}\left(  \eta\right)  \overset{\text{def}}{=}\sin\left(  \eta\right)
\arctan\left[  \mathrm{c}_{\mathrm{FS}}\tan\left(  \eta\right)  \right]  $, we
integrate by parts the integral in Eq. (\ref{cifro}) and get%
\begin{equation}
\int_{0}^{\tau}\eta\sin\left(  \eta\right)  \arctan\left[  \mathrm{c}%
_{\mathrm{FS}}\tan\left(  \eta\right)  \right]  d\eta=\left[  \eta
I_{V_{\mathrm{FS}}}\left(  \eta\right)  \right]  _{\eta=0}^{\eta=\tau}%
-\int_{0}^{\tau}I_{V_{\mathrm{FS}}}\left(  \eta\right)  d\eta\text{.}
\label{okyou}%
\end{equation}
Note that $g\left(  \eta\right)  =I_{V_{\mathrm{FS}}}\left(  \eta\right)  $
with $I_{V_{\mathrm{FS}}}\left(  \eta\right)  $ given in Eq. (\ref{funny}).
Finally, substituting Eq. (\ref{okyou}) into Eq. (\ref{cifro}) and considering
the asymptotic temporal behavior of $\mathcal{C}_{\mathrm{Sj\ddot{o}qvist}%
}\left(  \tau\right)  $, we obtain%
\begin{equation}
\mathcal{C}_{\mathrm{Sj\ddot{o}qvist}}^{\mathrm{asymptotic}}\left(
\tau\right)  =\frac{1}{8}\mathrm{a}_{\mathrm{FS}}\frac{\dot{r}_{i}}%
{\sqrt{1-r_{i}^{2}}}I_{V_{\mathrm{FS}}}^{\mathrm{asymptotic}}\left(
\tau\right)  \text{.} \label{com2}%
\end{equation}
At this point, considering the ratio between $\mathcal{C}_{\mathrm{Sj\ddot
{o}qvist}}^{\mathrm{asymptotic}}\left(  \tau\right)  $ in Eq. (\ref{com2}) and
the asymptotic temporal behavior of $\mathcal{C}_{\mathrm{FS}}\left(
\tau\right)  $ in Eq. (\ref{com1}), we get that the relative asymptotic
complexity growth in terms of a ratio exhibits a linear behavior given by
\begin{equation}
\frac{\mathcal{C}_{\mathrm{Sj\ddot{o}qvist}}^{\mathrm{asymptotic}}\left(
\tau\right)  }{\mathcal{C}_{\mathrm{FS}}^{\mathrm{asymptotic}}\left(
\tau\right)  }\sim\tau\text{.} \label{main1}%
\end{equation}
In the long-time limit, Eq. (\ref{main1}) expresses the fact that the
evolution of mixed states in the Bloch ball equipped with the Sj\"{o}qvist
metric explores averaged volumes of regions larger that the ones inspected
during the evolution of pure states on the Bloch sphere supplied with the
Fubini-Study metric. In particular, there appears to be an asymptotic linear
growth of the ratio between the two IGCs. Finally, in terms of the IGE defined
in Eq. (\ref{IGE}), we obtain an asymptotic entropy growth of the relative
difference between the two IGEs given by%
\begin{equation}
\mathcal{S}_{\mathrm{Sj\ddot{o}qvist}}^{\mathrm{asymptotic}}\left(
\tau\right)  -\mathcal{S}_{\mathrm{FS}}^{\mathrm{asymptotic}}\left(
\tau\right)  \sim\log\left(  \tau\right)  \text{.} \label{main2}%
\end{equation}
Eqs. (\ref{main2}) displays the asymptotic logarithmic discrepancy between the
IGE in the mixed and pure quantum state scenarios. Since the IGC is simply the
exponential of the IGE, we can interpret this entropic deviation as follows.
To a larger IGE there corresponds a larger IGC. Larger IGCs are larger
asymptotic averaged explored volumes. Larger volumes encode, via the metric,
larger fluctuations. The larger the fluctuations, the closer the points (i.e.,
the states) are together. The closer points are together, the greater is the
likelihood of incorrectly distinguishing quantum states during the evolutions
of the quantum system. This, in turn, leads to higher entropic configurations
which are typical of quantum systems in a mixed quantum state. Note that Eqs.
(\ref{ine1}), (\ref{compare}), (\ref{main1}), and (\ref{main2}) are
non-conflicting and consistent relations in support of arguments yielding to a
higher degree of complexity of evolutions of mixed states compared to pure
quantum states from a geometric perspective. Indeed, Eq. (\ref{ine1}) is an
inequality in terms of lengths. Eq. (\ref{compare}) is an inequality between
ratios expressed by means of accessible and instantaneous explored volumes.
Finally, Eqs. (\ref{main1}) and (\ref{main2}) are complexity and entropic
relations that are expressed by means of long-time limits of averaged explored
volumes of regions on and inside the Bloch ball. Interestingly, note that the
asymptotic temporal rates of change of the two IGCs in Eq. (\ref{main1}) scale
in a similar fashion, $d\mathcal{C}_{\mathrm{Sj\ddot{o}qvist}}%
^{\mathrm{asymptotic}}/d\tau\sim d\mathcal{C}_{\mathrm{FS}}%
^{\mathrm{asymptotic}}/d\tau$. This is a consequence of a balancing effect
that occurs between the asymptotic averaged explored \ volumes, $\bar
{V}_{\mathrm{Sj\ddot{o}qvist}}^{\mathrm{asymptotic}}\sim\tau\bar
{V}_{\mathrm{FS}}^{\mathrm{asymptotic}}$, and between the asymptotic temporal
rates of change of the averaged explored volumes, $d\bar{V}_{FS}%
^{\mathrm{asymptotic}}/d\tau\sim(1/\tau)d\bar{V}_{\mathrm{Sj\ddot{o}qvist}%
}^{\mathrm{asymptotic}}/d\tau$. From a curvature analysis perspective, the
manifold of pure states equipped with the Fubini-Study metric is an isotropic
two-dimensional manifold of constant positive sectional curvature
$\mathcal{K}_{\mathrm{FS}}=4$ and constant scalar curvature $\mathcal{R}%
_{\mathrm{FS}}=8$. Instead, the manifold of mixed states equipped with the
Sj\"{o}qvist metric is an anisotropic three-dimensional manifold of
non-constant but positive sectional curvature and constant scalar curvature
$\mathcal{R}_{\mathrm{Sj\ddot{o}qvist}}=8$. The positivity of sectional
curvatures in both scenarios leads to the presence of convergence in the
geodesic spread analysis on both manifolds. However, given the anisotropic
nature in the mixed quantum states manifold with the Sj\"{o}qvist metric, the
study of the geodesic spread equation would be more complicated in the
scenario of mixed states distinguished via the Sj\"{o}qvist metric. Our
remarks concerning the asymptotic rates of change of volumes and IGEs are an
indication of this distinct convergent behavior in the pure and mixed quantum
states scenarios. For further details on curvature properties along with a
comparison between the Sj\"{o}qvist and Bures manifolds, we refer to Appendix
E.\begin{table}[t]
\centering
\begin{tabular}
[c]{c|c|c|c|c}\hline\hline
\textbf{Type of state} & \textbf{Metric} & \textbf{Sectional curvature} &
\textbf{Path length} & \textbf{Information geometric complexity}\\\hline\hline
Pure & Fubini-Study & Constant & Shorter & Lower\\
Mixed & Sj\"{o}qvist & Nonconstant\qquad & Longer & Higher\\\hline
\end{tabular}
\caption{Schematic description of geometric properties (i.e., curvature,
length, and complexity) along evolutions of pure and mixed quantum states on
manifolds equipped with the Fubini-Study and Sj\"{o}qvist metrics,
respectively.}%
\end{table}

\section{Physical considerations}

In this section, we present physical comments on the concepts of metric, path
length, and curvature employed in our investigation. Moreover, we clarify the
physics behind the evolution of quantum states on curved manifolds in terms of
Bloch coordinates. These comments will help highlighting even further the
physical significance of our proposed complexity measure in Eq. (\ref{IGC}).

\subsection{Metric, path length, and curvature}

For completeness, we begin by recalling that in quantum mechanics a physical
state is not represented by a normalized state vector $\left\vert \psi\left(
t\right)  \right\rangle \in\mathcal{H}\backslash\left\{  0\right\}  $ but by a
ray. A ray is the one-dimensional subspace to which this vector belongs. Two
normalized vectors are equivalent, $\left\vert \psi^{\prime}\right\rangle
\sim$ $\left\vert \psi\right\rangle $, if they belong to the same ray, i.e.,
if $\left\vert \psi^{\prime}\right\rangle =e^{i\phi}\left\vert \psi
\right\rangle $ with $\phi\in U\left(  1\right)  $. This equivalence relation
specifies equivalence classes on the sphere $\mathcal{S}^{2N_{\mathcal{H}}+1}%
$, with $\dim_{%
%TCIMACRO{\U{2102} }%
%BeginExpansion
\mathbb{C}
%EndExpansion
}\mathcal{H}=N_{\mathcal{H}}+1$. Finally, the set of equivalence classes
$\mathcal{S}^{2N_{\mathcal{H}}+1}/U(1)$ forms the space of physical states
(rays) which is denoted here by $\mathcal{P}$. The space of rays is the
projective Hilbert space $\mathcal{P}$ which, in turn, is isomorphic to the
complex projective space $%
%TCIMACRO{\U{2102} }%
%BeginExpansion
\mathbb{C}
%EndExpansion
P^{N}$.

\subsubsection{Metric}

The metric (Eqs. (\ref{g4b}), (\ref{g11}), and (\ref{g12})) on the manifold of
quantum states is fixed once the quantum mechanical fluctuation in energy is
specified \cite{provost80}. \ Focusing on pure states, when the uncertainty
$\Delta A\left(  \eta\right)  $ in the generator of motion $A\left(
\eta\right)  $ with respect to the parameter $\eta$ in the projective Hilbert
space $\mathcal{P}$ is provided, the metric
\begin{equation}
ds_{\mathrm{FS}}^{2}=g_{\mu\nu}\left(  \xi\right)  \frac{d\xi^{\mu}}{d\eta
}\frac{d\xi^{\nu}}{d\eta}d\eta^{2}=\Delta A^{2}\left(  \eta\right)  d\eta^{2}
\label{phil}%
\end{equation}
is fixed. In particular, assuming $\eta=t$ and $A\left(  \eta\right)
=$\textrm{H}$\left(  t\right)  /\hslash$, we have that $ds_{\mathrm{FS}%
}=\left[  \Delta E\left(  t\right)  /\hslash\right]  dt$ and the dispersion
$\Delta E\left(  t\right)  $ of the generator of motion can originate from a
variety of Hamiltonians \textrm{H}$\left(  t\right)  $. Therefore, the
geometry of the projective Hilbert space, specified by the metric on it,
cannot be modified by the dynamics of the system governed by the Hamiltonian
\textrm{H}$\left(  t\right)  $ \cite{pati91}. This was a major result obtained
by Anandan and Aharonov in Ref. \cite{anandan90}. For pure states, the
distance function in the projective Hilbert space $\mathcal{P}$ is the
distance between two quantum states along a given curve in $\mathcal{P}$ as
measured by the Fubini-Study metric defined from the inner product of the
representative states in the $\left(  N_{\mathcal{H}}+1\right)  $-dimensional
Hilbert space $\mathcal{H}$. Anandan and Aharonov showed that this distance
equals the time integral of the uncertainty of the energy, and does not depend
on the particular Hamiltonian used to move the quantum system along a given
curve in $\mathcal{P}$. It is dependent only on the points in $\mathcal{P}$ to
which the quantum states project. In summary, from a physics standpoint, the
metric tensor and its components on the projective Hilbert space $\mathcal{P}$
are linked to the dispersion of suitable quantum-mechanical operators (for
instance, the Hamiltonian operator) acting on the underling Hilbert space
$\mathcal{H}$. This connection between metrics and quantum fluctuations is an
important physical consideration to keep in mind throughout our work. This
connection extends to the geometric analysis of quantum mixed states as well
\cite{zanardi07A,zanardi07B,erik20}.

\subsubsection{Path length}

To explain the physical meaning of the Riemannian distance (Eqs. (\ref{lf1})
and (\ref{sl1})) between two arbitrarily chosen pure quantum states, we follow
Wootters \cite{wootters81}. For mixed states, we hint to the work by
Braunstein and Caves in Ref. \cite{braunstein94}. Two infinitesimally close
points $\xi$ and $\xi+d\xi$ along a path $\xi\left(  \eta\right)  $ with
$\eta_{1}\leq\eta\leq\eta_{2}$ are statistically distinguishable if $d\xi$ is
at least equal to the standard fluctuation of $\xi$ \cite{diosi84}. The line
element along the path is $ds_{\mathrm{FS}}$ with $ds_{\mathrm{FS}}^{2}%
=g_{\mu\nu}^{\mathrm{FS}}\left(  \xi\right)  d\xi^{\mu}d\xi^{\nu}$. The length
of the path $\xi\left(  \eta\right)  $ with $\eta_{1}\leq\eta\leq\eta_{2}$
between $\xi_{1}\overset{\text{def}}{=}\xi\left(  \eta_{1}\right)  $ and
$\xi_{2}\overset{\text{def}}{=}\xi\left(  \eta_{2}\right)  $ is defined as%
\begin{equation}
\mathcal{L}\overset{\text{def}}{=}\int_{\xi_{1}}^{\xi_{2}}\sqrt
{ds_{\mathrm{FS}}^{2}}=\int_{\eta_{1}}^{\eta_{2}}\sqrt{\frac{ds_{\mathrm{FS}%
}^{2}}{d\eta^{2}}}d\eta\text{,}%
\end{equation}
and represents the maximal number $\tilde{N}$ of statistically distinguishable
states along the path. In particular, the geodesic distance between $\xi_{1}$
and $\xi_{2}$ is the path of shortest distance between $\xi_{1}$ and $\xi_{2}$
and is the minimum of $\tilde{N}$. This connection between path length and
number of statistically distinguishable states along the path is a relevant
physical remark to consider throughout our investigation. This viewpoint
extends naturally to the geometric analysis of quantum mixed states as well
\cite{braunstein94}.

\subsubsection{Curvature}

What is the physical significance of curvature (Appendix E) in our
investigation? We recall that in the Riemannian geometrization of classical
Newtonian mechanics \cite{pettini00}, the curvature $\mathcal{R}$ of the
manifold corresponds, roughly speaking, to the curvature of the potential $V$
expressed by means of the second derivative of $V$, $\mathcal{R}\sim
\partial^{2}V$, with the Hamiltonian of the system given by \textrm{H}$\left(
p\text{, }q\right)  =p^{2}/(2m)+V\left(  q\right)  $. More generally, in
arbitrary differential geometric settings, the curvature of the manifold
determines the stability (or, alternatively, the instability) of the geodesics
via the Jacobi equation of geodesic spread. This latter curvature
interpretation remains valid in our work. However, we can provide a more
specific interpretation for the concept of curvature in our analysis. As
pointed out by Braunstein and Caves in Ref. \cite{caves95}, unlike what
happens in general relativity, the geometry on the space of quantum states
does not describe the dynamical evolution of the physical system. Rather, it
places limits on our ability to discriminate one state from another via
measurements. In a sense, the geometry of quantum states puts the emphasis on
the fact that quantum mechanics is rooted in making statistical inferences
based on observed experimental data. Quantum measurement theory, in turn, is
statistical inference in its essence \cite{luo03}. Therefore, given the fact
that the problem of distinguishing neighboring quantum states can be
formulated as a parameter estimation problem \cite{braunstein94}, given that
quantum mechanics can be regarded as a theory for making statistical
inferences based on observed experimental data \cite{luo03}, and, finally,
since the curvature of a manifold is a measure of how difficult is to do
estimations at a given point in statistical science \cite{efron75}, it is
reasonable to interpret the curvature of a manifold of quantum states equipped
with suitably defined metric structures as an indicator of how difficult is to
distinguish quantum states by means of parameter estimation at a given point
of the state space. In particular, the higher the curvature, the more
difficult is to do estimation at that point. This is the point of view that we
adopt in this paper. For remarks on a physical interpretation of curvature of
manifolds underlying the information geometry of non-interacting gases
satisfying the Fermi-Dirac and Bose-Einstein statistics, we refer to Ref.
\cite{pessoa21}. Finally, for a work on the estimation of the curvature of a
quantum manifold via measurement on a quantum particle constrained to
propagate on the manifold itself, we hint to Ref. \cite{paris19}.

\subsection{Evolution of Bloch coordinates}

Having clarified the meaning of metric, path length, and curvature employed in
our work we devote some time explaining the relation between Bloch coordinates
and quantum states on the Bloch sphere and inside the Bloch ball (Appendix A,
C, and D). Let us point out from the start that in our work $\xi^{\mu}\left(
\eta\right)  \overset{\text{def}}{=}\left(  \xi^{1}\left(  \eta\right)
\text{, }\xi^{2}\left(  \eta\right)  \text{, }\xi^{3}\left(  \eta\right)
\right)  $ in Eq. (\ref{phil}) is specified by the Bloch coordinates, that is,
$\xi^{\mu}\left(  \eta\right)  =\left(  r\left(  \eta\right)  \text{, }%
\theta\left(  \eta\right)  \text{, }\varphi\left(  \eta\right)  \right)  $.
\ In this paper, we focused on the integration of evolution equations for
Bloch coordinates used to parametrize quantum states, either pure or mixed.
This choice was not dictated by mathematical convenience only. Indeed, there
is a clear physical path connecting Bloch coordinates, Bloch vectors, and,
finally, pure and mixed quantum states. For simplicity, we set $\eta=t$ in
this discussion. We observe that from the time evolution of the Bloch
coordinates, both radial and angular, one can generally recover the time
evolution of the density operators for arbitrary quantum states via the time
evolution of the Bloch vector. Conversely, the opposite is also possible. For
a detailed study concerning the time evolution of the Bloch vector of a single
two-level atom that interacts with a single quantized electromagnetic field
mode according to the Jaynes-Cummings model, we refer to Ref. \cite{azuma08}.
To be explicit here, we note that the Bloch vector $\vec{p}\left(  t\right)  $
is defined as%
\begin{equation}
\vec{p}\left(  t\right)  \overset{\text{def}}{=}\mathrm{tr}\left[  \rho\left(
t\right)  \vec{\sigma}\right]  =\left(  r\sin\theta\cos\varphi\text{, }%
r\sin\theta\sin\varphi\text{, }r\cos\theta\right)  \text{,} \label{bloch}%
\end{equation}
with $r=r\left(  t\right)  $, $\theta=\theta\left(  t\right)  $, and
$\varphi=\varphi\left(  t\right)  $. Focusing for simplicity on the case of
unitary quantum evolution, the density operator $\rho\left(  t\right)
=(1/2)\left[  \mathrm{I}+\vec{p}\left(  t\right)  \cdot\vec{\sigma}\right]  $
in Eq. (\ref{bloch}) satisfies the von Neumann equation $i\hslash\dot{\rho
}=\left[  \mathrm{H}\left(  t\right)  \text{, }\rho\left(  t\right)  \right]
$ with $\dot{\rho}\overset{\text{def}}{=}d\rho/dt$ and $\mathrm{I}$ denoting
the identity operator on the single-qubit quantum state space. Moreover, for a
system in a pure state that evolves under a time independent Hamiltonian
\textrm{H}, $\rho\left(  t\right)  =\left\vert \psi\left(  t\right)
\right\rangle \left\langle \psi\left(  t\right)  \right\vert =U\left(
t\right)  \rho\left(  0\right)  U^{\dagger}\left(  t\right)  $ with $U\left(
t\right)  $ being the unitary time evolution operator given by $\exp\left[
-(i/\hslash)\mathrm{H}t\right]  $ and $\left\vert \psi\left(  t\right)
\right\rangle =\cos\left[  \theta\left(  t\right)  /2\right]  \left\vert
0\right\rangle +e^{i\varphi\left(  t\right)  }\sin\left[  \theta\left(
t\right)  /2\right]  \left\vert 1\right\rangle $ in the Bloch sphere
parametrization. To make very clear this link among Bloch coordinates, Bloch
vectors, pure states, and mixed states, we consider a simple illustrative
example for pure states evolution. Assume the system, a spin-$1/2$ particle,
is initially in the state $\left\vert \psi\left(  0\right)  \right\rangle $
parametrized in terms of $\left(  \theta\left(  0\right)  \text{, }%
\varphi\left(  0\right)  \right)  =\left(  \pi/2\text{, }0\right)  $. Solving
the Schr\"{o}dinger's evolution equation $i\hslash\partial_{t}\left\vert
\psi\left(  t\right)  \right\rangle =$\textrm{H}$\left(  t\right)  \left\vert
\psi\left(  t\right)  \right\rangle $ with \textrm{H}$\overset{\text{def}%
}{=}\hslash\omega_{0}\sigma_{z}$, we get%
\begin{equation}
\rho\left(  t\right)  =\left\vert \psi\left(  t\right)  \right\rangle
\left\langle \psi\left(  t\right)  \right\vert =\frac{1}{2}\left(
\begin{array}
[c]{cc}%
1 & e^{-2i\omega_{o}t}\\
e^{2i\omega_{o}t} & 1
\end{array}
\right)  =\frac{1}{2}\left[  \mathrm{I}+\vec{p}\left(  t\right)  \cdot
\vec{\sigma}\right]  \text{,}%
\end{equation}
with $\vec{p}\left(  t\right)  =\left(  \cos\left(  2\omega_{0}t\right)
\text{, }\sin\left(  2\omega_{0}t\right)  \text{, }0\right)  $. Furthermore,
the Bloch angles at time $t$ become $\left(  \theta\left(  t\right)  \text{,
}\varphi\left(  t\right)  \right)  =\left(  \pi/2\text{, }2\omega_{0}t\right)
$. In general, the state $\left\vert \psi\left(  t\right)  \right\rangle
=U\left(  t\right)  \left\vert \psi\left(  0\right)  \right\rangle $ can be
parametrized as $\left\vert \psi\left(  t\right)  \right\rangle =a\left(
t\right)  \left\vert 0\right\rangle +b\left(  t\right)  \left\vert
1\right\rangle $ with $%
%TCIMACRO{\U{2102} }%
%BeginExpansion
\mathbb{C}
%EndExpansion
\ni a\left(  t\right)  =\left\vert a\left(  t\right)  \right\vert
e^{i\varphi_{a}\left(  t\right)  }$, $%
%TCIMACRO{\U{2102} }%
%BeginExpansion
\mathbb{C}
%EndExpansion
\ni b\left(  t\right)  =\left\vert b\left(  t\right)  \right\vert
e^{i\varphi_{b}\left(  t\right)  }$, $\varphi_{a\text{, }b}\left(  t\right)
\in%
%TCIMACRO{\U{211d} }%
%BeginExpansion
\mathbb{R}
%EndExpansion
$, and $\left\vert a\left(  t\right)  \right\vert ^{2}+\left\vert b\left(
t\right)  \right\vert ^{2}=1$. With this pure state parametrization, the Bloch
angles $\theta\left(  t\right)  $ and $\varphi\left(  t\right)  $ satisfy the
relations, $\tan\left[  \theta\left(  t\right)  \right]  =\left\vert b\left(
t\right)  \right\vert /\left\vert a\left(  t\right)  \right\vert $ and
$\varphi\left(  t\right)  =\varphi_{b}\left(  t\right)  -\varphi_{a}\left(
t\right)  $, respectively.

Having clarified the physical meaning of geometrical concepts employed in our
analysis, the main take-home message is the following. We have estimated in
this paper the complexity of geodesic paths of both pure and mixed quantum
states by means of a complexity measure (Eq. (\ref{IGC})) expressed in terms
of explored volumes of the suitably metricized curved manifolds that underlay
the dynamics (i.e., the change in Bloch parameters, with changes specified by
the parametric evolution operator). The metric structure on the curved
manifolds of quantum states is fixed by quantum-mechanical fluctuations.
Moreover, just as path lengths can be interpreted in terms of the maximal
number of distinguishable states traversed during the evolution along the
path, the volumes of the parametric space explored in a fixed temporal
interval can be regarded as representing the maximal number of different
states visited during the regional exploration. Clearly, the role played by
the infinitesimal increment $d\xi$ in the path exploration is replaced by the
infinitesimal volume element $dV\overset{\text{def}}{=}\sqrt{g\left(
\xi\right)  }d^{N}\xi$ in the regional travel, with $g\left(  \xi\right)
\overset{\text{def}}{=}\det\left[  g_{\mu\nu}\left(  \xi\right)  \right]  $
and $N$ being the dimensionality of the curved manifold. Essentially, the
Riemannian volume element $dV$ helps gauging the number of distinct states
explored within an infinitesimal volume of a region of the manifold
\cite{myung00}. We are ready now for our conclusions.

\section{Concluding Remarks}

We present here a summary of our main findings along with limitations and
possible future directions.

\subsection{Summary of results}

In this paper, we provided a comparative information geometric analysis of the
complexity of geodesic paths of pure and mixed quantum states on the Bloch
sphere and inside the Bloch ball, respectively. In this geometric setting,
pure and mixed states were chosen to be distinguished by means of the
Fubini-Study (Eq. (\ref{g4b})) and the Sj\"{o}qvist metric (Eq. (\ref{g11})),
respectively. After finding the geodesic paths connecting arbitrary points on
(see Appendix A) and inside the Bloch ball (see Appendix C), we analytically
estimated the IGE (Eq. (\ref{IGE})) and the IGC (Eq. (\ref{IGC})) in both
scenarios. The long-time limit of this pair of entropic measures of complexity
of evolution of system in pure (see Eq. (\ref{com1})) and mixed (see Eq.
(\ref{com2})) states were compared. We observed a degree of complexity for the
evolution of mixed states with the Sj\"{o}qvist geometry higher than the one
specifying the complexity for the evolution of pure states with the
Fubini-Study geometry (see Eqs. (\ref{main1}) and (\ref{main2})).

The metric structure on the manifold of quantum states is specified by quantum
fluctuations. Path lengths and volumes can be physically interpreted as
indicators of the maximal number of distinguishable states crossed along
trajectories and in volumes of regions of the manifold, respectively. To a
higher count of distinct states passed over in a fixed time interval, there
corresponds a higher degree of complexity of the evolution on the underlying
manifold. Within this physically meaningful geometric description, mixed state
(geodesic) evolutions appear to be generally more complex than pure state evolutions.

Our main findings can be outlined as follows:

\begin{enumerate}
\item[{[1]}] We proposed a different information geometric way (Eqs.
(\ref{IGE}) and (\ref{IGC})) to describe and, to a certain extent, understand
the complex behavior of evolutions of quantum systems in pure and mixed
states. The ranking is probabilistic in nature, it requires a temporal
averaging procedure along with a long-time limit, and is limited to comparing
expected geodesic evolutions on the underlying manifolds.

\item[{[2]}] We showed (Eqs. (\ref{main1}) and (\ref{main2})) that the
complexity of geodesic paths (Appendix C) corresponding to the evolution of
mixed quantum states in the Bloch ball equipped with the Sj\"{o}qvist metric
is higher than the complexity of geodesic paths (Appendix A) arising from the
evolution of pure states on the Bloch sphere furnished with the Fubini-Study metric.

\item[{[3]}] We found that the ranking in terms of the information geometric
complexity (\ref{main1}), a quantity that represents the asymptotic temporal
behavior of an averaged volume of the region explored on the manifold during
the evolution, is in agreement with the ranking in terms of lengths (Eq.
(\ref{ine1})) and, in addition, volume ratios in terms of accessible and
instantaneous explored volumes (\ref{compare}). For a schematic summary, we
refer to Table II.

\item[{[4]}] We \ confirmed that the choice of the metric on the space of
mixed states matters. Specifically, we observed fingerprints of a softening of
the complexity on the Bures manifold (Eq. (\ref{boys})) compared to the
Sj\"{o}qvist manifold. This is in agreement with the presence of longer
lengths of geodesic paths on the Sj\"{o}qvist manifold (Eq. (\ref{amo2})).
Furthermore, the two manifolds exhibit different curvature properties. The
Bures manifold is isotropic, while the Sj\"{o}qvist manifold is anisotropic
(Appendix E). For a schematic outline, we hint to Table III.\begin{table}[t]
\centering
\begin{tabular}
[c]{c|c|c|c}\hline\hline
\textbf{Metric} & \textbf{Sectional curvature} & \textbf{Path length} &
\textbf{Information geometric complexity}\\\hline\hline
Sj\"{o}qvist & Nonconstant & Longer & Stronger\\
Bures & Constant & Shorter & Weaker\\\hline
\end{tabular}
\caption{Schematic description of distinct features of the Sj\"{o}qvist and
Bures metrics in terms of sectional curvatures, path lengths, and information
geometric complexities.}%
\end{table}
\end{enumerate}

\subsection{Outlook}

Like most scientific studies, our investigation suffers a few limitations.
From a computational standpoint, our proposed complexity measure requires
volume calculations which are much harder than action or path length
calculations since there are differential equations to be solved. This
particular point is in agreement with what stated in Ref. \cite{leo16B}.
Therefore, exact analytical solutions are rare and approximate numerical
solutions are unavoidable in more realistic physical scenarios. From a
conceptual perspective, there are at least two weaknesses. First, there is a
freedom in the choice of the metric for mixed quantum states. For further
details on the Sj\"{o}qvist metric that concern its extension to the
degenerate case along with its relation to the Bures metric, we refer to
Appendix F. Second, we only limited our work to the study of two-level quantum
systems. The ambiguity in the metric affects the notion of speed which, in
turn, is related to the concepts of length, action, and complexity. The
dependence of the ratio of distance and time on the choice of the metric on
the space of mixed quantum states is in agreement with the considerations
carried out in Ref. \cite{brody19}. The restriction to a single qubit, while
simple and insightful, cannot be expected to cover the full richness of a
higher-dimensional quantum dynamics occurring in an exponentially larger
Hilbert space. In particular, scaling laws with respect to the dimensionality
of the Hilbert space of suitable physical quantities cannot be addressed in
this limiting scenario. Moreover, one fundamental quantum phenomenon that
escapes a single qubit treatment is entanglement. This second limitation is
similar to the one presented in Ref. \cite{brown19}.

Despite these limitations, we believe our work is relevant also in view of the
fact that it paves the way to further lines of inquiry. For instance, it
naturally triggers the following questions: 1) Using our findings along with
the ones presented in Refs. \cite{nielsen06,brown19}, can one compare the
complexity of geodesic motion on differently deformed Bloch spheres by adding
anisotropic penalty factors to the Fubini-Study metric?; 2) What happens to
the evolution of mixed states inside a deformed Bloch ball when introducing
anisotropic penalty factors? Is the relative ranking in terms of complexity
between pure and mixed states preserved under any arbitrary deformations of
the Bloch sphere?; 3) Can one compare the complexity of geodesic paths on a
deformed Bloch sphere with the complexity of geodesic paths in a non deformed
Bloch ball? 4) Is there a minimal complexity metric for mixed quantum states?;
5) Using our curvature calculations and the analysis presented in Ref.
\cite{auzzi21}, how much does one need to deform the Bloch sphere ( that is,
introducing anisotropic penalty factors to get negative sectional curvature)
and how high should the dimensionality of a quantum system be in order to
address the issue of ergodicity and properly apply thermodynamic arguments to
complexity evolution? The introduction of anisotropic penalty factors can be
motivated by experimental considerations. For example, considering a
spin-$1/2$ particle in an external magnetic field, it may be the case that is
is easier to apply the field in some direction rather that in another. In this
case, the penalty factor would be larger where it is more complicated to apply
the field. Incorporating these factors in our analysis would open up to lines
of investigation of relevance in the context of finding optimal Hamiltonian
evolutions, both in terms of efficiency
\cite{cafaropra20,cafaropra22,ijqi19,ps19,steven20,f19,e21} and complexity
\cite{cafaropre20,cafaropre22}, in the presence of physical constraints
dictated by experimental limitations.

We hope our work will inspire other scientists and pave the way toward further
investigations in this fascinating research direction. For the time being, we
leave a more in-depth quantitative discussion on these potential extensions
and applications of our theoretical findings to future scientific
efforts.\bigskip

\begin{acknowledgments}
C.C. is grateful to the United States Air Force Research Laboratory (AFRL)
Summer Faculty Fellowship Program for providing support for this work. P.M.A.
acknowledges support from the Air Force Office of Scientific Research (AFOSR).
Any opinions, findings and conclusions or recommendations expressed in this
material are those of the author(s) and do not necessarily reflect the views
of the Air Force Research Laboratory (AFRL).
\end{acknowledgments}

\bigskip\pagebreak

\appendix

\section{Geodesic paths on the Bloch sphere}

In this Appendix, we derive the geodesic paths on the two-sphere. In the first
derivation, we use simple geometric arguments to obtain the equation of a
great circle in spherical coordinates. In the second derivation, we integrate
the geodesic equations to get explicit expressions of geodesic paths
$\theta=\theta\left(  \eta\right)  $ and $\varphi=\varphi\left(  \eta\right)
$. Then, combining the geodesic paths equations, we show that we also get the
equation of a great circle in spherical coordinates that matches our first derivation.

\subsection{Geometric derivation}

Geodesics on the two-sphere lie on great circles. Great circles, in turn, can
be obtained by intersecting a plane passing through the origin in $%
%TCIMACRO{\U{211d} }%
%BeginExpansion
\mathbb{R}
%EndExpansion
^{3}$ with the two-sphere. Assume that the equations of the plane and the
two-sphere of unit radius are given by,%
\begin{equation}
\alpha x+\beta y+\gamma z=0\text{, and }x^{2}+y^{2}+z^{2}=1\text{,}
\label{A21}%
\end{equation}
respectively. In Eq. (\ref{A21}), $\alpha$, $\beta$, and $\gamma$ belong to $%
%TCIMACRO{\U{211d} }%
%BeginExpansion
\mathbb{R}
%EndExpansion
$. Using spherical coordinates, we set $x\overset{\text{def}}{=}\sin\left(
\theta\right)  \cos\left(  \varphi\right)  $, $y\overset{\text{def}}{=}%
\sin\left(  \theta\right)  \sin\left(  \varphi\right)  $, and
$z\overset{\text{def}}{=}\cos\left(  \theta\right)  $. Finally, combining the
two relations in Eq. (\ref{A21}), we get the equation of a great circle in
spherical coordinates%
\begin{equation}
\cot\left(  \theta\right)  =\pm a\cos\left(  \varphi-\bar{\varphi}\right)
\text{.} \label{A22}%
\end{equation}
The constants $a$ and $\bar{\varphi}$ in Eq. (\ref{A22}) are such that
$a^{2}\overset{\text{def}}{=}\left(  \alpha^{2}+\beta^{2}\right)  /\gamma^{2}$
and $\tan\left(  \bar{\varphi}\right)  \overset{\text{def}}{=}\beta/\alpha$, respectively.

\subsection{Dynamics derivation}

Consider the system of two coupled second order nonlinear ODEs,%
\begin{equation}
\left\{
\begin{array}
[c]{c}%
\frac{d^{2}\theta}{d\eta^{2}}-\sin\left(  \theta\right)  \cos\left(
\theta\right)  \left(  \frac{d\varphi}{d\eta}\right)  ^{2}=0\\
\frac{d^{2}\varphi}{d\eta^{2}}+2\frac{\cos\left(  \theta\right)  }{\sin\left(
\theta\right)  }\frac{d\theta}{d\eta}\frac{d\varphi}{d\eta}=0
\end{array}
\right.  \text{.} \label{A23}%
\end{equation}
Note that the second relation in Eq. (\ref{A23}) is equivalent to,%
\begin{equation}
\frac{d}{d\eta}\left(  \frac{d\varphi}{d\eta}\sin^{2}\left(  \theta\right)
\right)  =0\text{,}%
\end{equation}
that is,%
\begin{equation}
\left(  \frac{d\varphi}{d\eta}\right)  ^{2}=\frac{\mathrm{c}_{\mathrm{FS}}%
^{2}}{\sin^{4}\left(  \theta\right)  }\text{,} \label{A24}%
\end{equation}
with $\mathrm{c}_{\mathrm{FS}}=\mathrm{c}_{\mathrm{FS}}\left(  \theta
_{i}\text{, }\dot{\varphi}_{i}\right)  \overset{\text{def}}{=}\dot{\varphi
}_{i}\sin^{2}\left(  \theta_{i}\right)  $ being a real constant where
$\theta_{i}\overset{\text{def}}{=}\theta\left(  \eta_{i}\right)  $ and
$\dot{\varphi}_{i}\overset{\text{def}}{=}(d\varphi/d\eta)_{\eta=\eta_{i}}$
with $\eta_{i}$ set equal to zero. Using Eq. (\ref{A24}), the first relation
in Eq. (\ref{A23}) yields%
\begin{equation}
\frac{d^{2}\theta}{d\eta^{2}}=\mathrm{c}_{\mathrm{FS}}^{2}\frac{\cos\left(
\theta\right)  }{\sin^{3}\left(  \theta\right)  }\text{.} \label{A25}%
\end{equation}
From Eqs. (\ref{A24}) and (\ref{A25}), we note that the original system of
coupled ODEs in Eq. (\ref{A23}) can be uncoupled. However, Eq. (\ref{A25}) is
nonlinear and we shall use some tricks to find a way of integrating it.
Imposing the unit-speed condition, $\dot{\theta}^{2}+\dot{\varphi}^{2}\sin
^{2}\left(  \theta\right)  =\mathrm{const.}$with $\mathrm{const.}=1$, we
employ Eqs. (\ref{A24}) and (\ref{A25}) to get%
\begin{equation}
\int d\eta=\pm\int\frac{\sin\left(  \theta\right)  }{\sqrt{\sin^{2}\left(
\theta\right)  -\mathrm{c}_{\mathrm{FS}}^{2}}}d\theta\text{.} \label{A26}%
\end{equation}
Let us perform a change of variables and put $\epsilon\overset{\text{def}%
}{=}\cos\left(  \theta\right)  $. Then, integration of Eq. (\ref{A26}) yields%
\begin{equation}
\eta=\eta_{i}\pm\left[  \arctan\left(  \frac{\epsilon_{i}}{\sqrt
{\mathrm{a}_{\mathrm{FS}}^{2}-\epsilon_{i}^{2}}}\right)  -\arctan\left(
\frac{\epsilon}{\sqrt{\mathrm{a}_{\mathrm{FS}}^{2}-\epsilon^{2}}}\right)
\right]  \text{,} \label{A27}%
\end{equation}
where $\mathrm{a}_{\mathrm{FS}}^{2}\overset{\text{def}}{=}1-\mathrm{c}%
_{\mathrm{FS}}^{2}$. For simplicity, assume $\theta_{i}\overset{\text{def}%
}{=}\theta\left(  \eta_{i}\right)  =\cos^{-1}\left(  \epsilon_{i}\right)
=\pi/2$ with $\eta_{i}=0$. Then, $\epsilon_{i}=0$ and manipulation of Eq.
(\ref{A27}) yields%
\begin{equation}
\cos\left(  \theta\right)  =\pm\mathrm{a}_{\mathrm{FS}}\sin\left(
\eta\right)  \text{.} \label{A28}%
\end{equation}
We remark that for an arbitrary $\theta_{i}$, the analogue of Eq. (\ref{A28})
squared is simply given by%
\begin{equation}
\cos^{2}\left(  \theta\right)  =\mathrm{a}_{\mathrm{FS}}^{2}\sin^{2}\left[
\eta+\arctan\left(  \frac{\epsilon_{i}}{\sqrt{\mathrm{a}_{\mathrm{FS}}%
^{2}-\epsilon_{i}^{2}}}\right)  \right]  \text{.} \label{A28B}%
\end{equation}
As a consistency check, we note that for $\eta_{i}=0$, Eq. (\ref{A28B})
correctly yields
\begin{equation}
\cos^{2}\left(  \theta_{i}\right)  =\mathrm{a}_{\mathrm{FS}}^{2}\sin
^{2}\left[  \arctan\left(  \frac{\epsilon_{i}}{\sqrt{\mathrm{a}_{\mathrm{FS}%
}^{2}-\epsilon_{i}^{2}}}\right)  \right]  =\epsilon_{i}^{2}\text{.}
\label{A28C}%
\end{equation}
We also point out here that we could have set $\tilde{v}_{\mathrm{FS}}%
^{2}\overset{\text{def}}{=}\dot{\theta}^{2}+\dot{\varphi}^{2}\sin^{2}\left(
\theta\right)  =\mathrm{const.}$with $\mathrm{const.}\neq1$. Then, Eq.
(\ref{A28}) would simply become $\cos\left(  \theta\right)  =\pm
\mathrm{a}_{\mathrm{FS}}\sin\left(  \tilde{v}_{\mathrm{FS}}\eta\right)  $ with
$\mathrm{a}_{\mathrm{FS}}^{2}\overset{\text{def}}{=}1-\left(  \mathrm{c}%
_{\mathrm{FS}}/\tilde{v}_{\mathrm{FS}}\right)  ^{2}$. Eq. (\ref{A28}) allows
us to express $\theta=\theta\left(  \eta\right)  $. Next, we need to find the
relation $\varphi=\varphi\left(  \eta\right)  $. Using Eqs. (\ref{A24}) and
(\ref{A28}), we get%
\begin{equation}
\frac{d\varphi}{d\eta}=\frac{\mathrm{c}_{\mathrm{FS}}}{1-\mathrm{a}%
_{\mathrm{FS}}^{2}\sin^{2}\left(  \eta\right)  }\text{.} \label{A29}%
\end{equation}
Integration of Eq. (\ref{A29}) leads to,%
\begin{equation}
\varphi\left(  \eta\right)  =\varphi_{i}+\frac{\mathrm{c}_{\mathrm{FS}}}%
{\sqrt{-1+\mathrm{a}_{\mathrm{FS}}^{2}}}\tanh^{-1}\left[  \sqrt{-1+\mathrm{a}%
_{\mathrm{FS}}^{2}}\tan\left(  \eta\right)  \right]  \text{.} \label{A30}%
\end{equation}
Note that $\sqrt{-1+\mathrm{a}_{\mathrm{FS}}^{2}}=i\mathrm{c}_{\mathrm{FS}}$,
with $i\in%
%TCIMACRO{\U{2102} }%
%BeginExpansion
\mathbb{C}
%EndExpansion
$ denoting the imaginary unit. Furthermore, recalling from Ref. \cite{irene}
that the inverse hyperbolic tangent function of a complex variable satisfies
the relation $\tanh^{-1}\left(  z\right)  =-i\tan^{-1}\left(  iz\right)  $,
setting $z\overset{\text{def}}{=}ix$ with $x\in%
%TCIMACRO{\U{211d} }%
%BeginExpansion
\mathbb{R}
%EndExpansion
$ and $z\in%
%TCIMACRO{\U{2102} }%
%BeginExpansion
\mathbb{C}
%EndExpansion
$, we obtain%
\begin{equation}
-i\tanh^{-1}\left(  ix\right)  =\arctan\left(  x\right)  \text{.} \label{A31}%
\end{equation}
Finally, using Eqs. (\ref{A30}) and (\ref{A31}), we get%
\begin{equation}
\varphi\left(  \eta\right)  =\varphi_{i}+\tan^{-1}\left[  \mathrm{c}%
_{\mathrm{FS}}\tan\left(  \eta\right)  \right]  \text{,}%
\end{equation}
that is,%
\begin{equation}
\tan\left[  \varphi\left(  \eta\right)  -\varphi_{i}\right]  =\mathrm{c}%
_{\mathrm{FS}}\tan\left(  \eta\right)  \text{.} \label{A32}%
\end{equation}
For completeness, we point out that using Eqs. (\ref{A28}) and (\ref{A32})
along with the following two trigonometric identities,%
\begin{equation}
\cos^{2}\left(  \theta\right)  =\frac{\cot^{2}\left(  \theta\right)  }%
{1+\cot^{2}\left(  \theta\right)  }\text{, and }\sin^{2}\left(  \eta\right)
=\frac{\tan^{2}\left(  \eta\right)  }{1+\tan^{2}\left(  \eta\right)  }\text{,}%
\end{equation}
we get after some straightforward algebraic manipulations the equation of a
great circle,%
\begin{equation}
\cot\left(  \theta\right)  =\pm\sqrt{\frac{1-\mathrm{c}_{\mathrm{FS}}^{2}%
}{\mathrm{c}_{\mathrm{FS}}^{2}}}\sin\left(  \varphi-\varphi_{i}\right)
\text{.} \label{A33}%
\end{equation}
Eq. (\ref{A33}) is identical to Eq. (\ref{A22}) once we identify $a$ and
$\bar{\varphi}$ with $\sqrt{\frac{1-\mathrm{c}_{\mathrm{FS}}^{2}}%
{\mathrm{c}_{\mathrm{FS}}^{2}}}$ and $\varphi_{i}+\pi/2$, respectively.

\section{The Bures infinitesimal line element}

In this Appendix, we derive Eq. (\ref{g12}). Recall that for the single qubit
case the Bures distance between two infinitesimally close density matrices
$\rho$ and $\rho+d\rho$ is given by \cite{karol03},%
\begin{equation}
ds_{\mathrm{Bures}}^{2}\overset{\text{def}}{=}\frac{1}{2}\sum_{n,m=0}^{1}%
\frac{\left\vert \left\langle e_{m}|d\rho|e_{n}\right\rangle \right\vert ^{2}%
}{p_{m}+p_{n}}\text{,} \label{ag1}%
\end{equation}
where $\left\{  \left\vert e_{n}\right\rangle \right\}  _{n=0,1}$ is an
orthonormal basis of eigenvectors of $\rho$ with eigenvalues $\left\{
p_{n}\right\}  _{n=0,1}$. Therefore, $\rho=p_{0}\left\vert e_{0}\right\rangle
\left\langle e_{0}\right\vert +p_{1}\left\vert e_{1}\right\rangle \left\langle
e_{1}\right\vert $. From the expression of $\rho$ in Eq. (\ref{g7}), we get%
\begin{equation}
d\rho=\frac{\partial\rho}{\partial r}dr+\frac{\partial\rho}{\partial\theta
}d\theta+\frac{\partial\rho}{\partial\varphi}d\varphi\text{,} \label{ag2}%
\end{equation}
that is,%
\begin{equation}
d\rho=\frac{1}{2}\left(
\begin{array}
[c]{cc}%
\cos\left(  \theta\right)  dr-r\sin\left(  \theta\right)  d\theta &
e^{-i\varphi}\left[  \sin\left(  \theta\right)  dr+r\cos\left(  \theta\right)
-ir\sin\left(  \theta\right)  d\varphi\right] \\
e^{i\varphi}\left[  \sin\left(  \theta\right)  dr+r\cos\left(  \theta\right)
+ir\sin\left(  \theta\right)  d\varphi\right]  & -\cos\left(  \theta\right)
dr+r\sin\left(  \theta\right)  d\theta
\end{array}
\right)  \text{.} \label{ag3}%
\end{equation}
Finally, using Eqs. (\ref{g8}), (\ref{g9}), (\ref{g10}), and (\ref{ag3}),
$ds_{\mathrm{Bures}}^{2}$ in Eq. (\ref{ag1}) reduces to $ds_{\mathrm{Bures}%
}^{2}$ in Eq. (\ref{g12}). For further details on the Bures metric for
high-dimensional quantum systems, we refer to Refs. \cite{slater98,karol}.

\section{Geodesic paths in the Bloch ball with the Sj\"{o}qvist metric}

In this Appendix, we study the geodesic paths in the Bloch ball using the
Sj\"{o}qvist metric. Recall that the form of the geodesic equation remains
unchanged under affine transformations. An affine transformation of a
parameter $\eta$ is the change of variable $\eta\rightarrow\eta^{\prime
}=\tilde{a}$ $\eta+\tilde{b}$, with $\tilde{a}$ and $\tilde{b}$ in $%
%TCIMACRO{\U{211d} }%
%BeginExpansion
\mathbb{R}
%EndExpansion
$. Any other transformation will generate extra terms in the geodesic equation
\cite{epj}. The affine parameter for a geodesic is unique up to an affine
change of variables. Therefore, it is important to keep in mind that geodesics
are curves with a preferred parametrization. In what follows, we explicitly
analyze geodesic paths parametrized in terms of two distinct affine parametrizations.

\subsection{The Sj\"{o}qvist $\theta$-affine parametrization}

In Ref. \cite{erik20}, Sj\"{o}qvist focused on finding geodesic paths
connecting points in the Bloch ball laying in a plane that contains the
origin. In other words, using spherical coordinates $\left(  r\text{, }%
\theta\text{, }\varphi\right)  $ and keeping $\varphi=$\textrm{const.},
geodesics were obtained by minimizing $\int$ $ds_{\mathrm{Sj\ddot{o}qvist}}$
over all curves connecting points $\left(  r_{i}\text{, }\theta_{i}\right)  $
and $\left(  r_{f}\text{, }\theta_{f}\right)  $. More specifically, one
obtains the curve $\left[  \theta_{i}\text{, }\theta_{f}\right]  \ni
\theta\mapsto r_{\mathrm{Sj\ddot{o}qvist}}\left(  \theta\right)  \in\left(
0\text{, }1\right]  $ that minimizes the length $\mathcal{L}_{\mathrm{Sj\ddot
{o}qvist}}\left(  \theta_{i}\text{, }\theta_{f}\right)  $ defined as%
\begin{equation}
\mathcal{L}_{\mathrm{Sj\ddot{o}qvist}}\left(  \theta_{i}\text{, }\theta
_{f}\right)  \overset{\text{def}}{=}\int_{s_{i}}^{s_{f}}\sqrt
{ds_{\mathrm{Sj\ddot{o}qvist}}^{2}}=\frac{1}{2}\int_{\theta_{i}}^{\theta_{f}%
}\mathrm{L}\left(  r^{\prime}\text{, }r\text{, }\theta\right)  d\theta\text{.}
\label{s1}%
\end{equation}
In Eq. (\ref{s1}), $r^{\prime}\overset{\text{def}}{=}dr/d\theta$ and
$\mathrm{L}\left(  r^{\prime}\text{, }r\text{, }\theta\right)  $ is the
Lagrangian-like function given by%
\begin{equation}
\mathrm{L}\left(  r^{\prime}\text{, }r\text{, }\theta\right)
\overset{\text{def}}{=}\sqrt{1+\frac{r^{\prime2}}{1-r^{2}}}\text{.} \label{s2}%
\end{equation}
From Eq. (\ref{s2}), note that $\mathrm{L}=\mathrm{L}\left(  r^{\prime}\text{,
}r\right)  $ does not depend explicitly on $\theta$. Therefore, $\partial
\mathrm{L}/\partial\theta=0$. In this case, it happens that the Euler-Lagrange
equation%
\begin{equation}
\frac{d}{d\theta}\frac{\partial\mathrm{L}\left(  r^{\prime}\text{, }r\right)
}{\partial r^{\prime}}-\frac{\partial\mathrm{L}\left(  r^{\prime}\text{,
}r\right)  }{\partial r}=0\text{,} \label{EL}%
\end{equation}
reduces to the so-called Beltrami identity,%
\begin{equation}
\mathrm{L}\left(  r^{\prime}\text{, }r\right)  -r^{\prime}\frac{\partial
\mathrm{L}\left(  r^{\prime}\text{, }r\right)  }{\partial r^{\prime}%
}=\text{\textrm{const}.} \label{Beltrami}%
\end{equation}
Indeed, combining Eq. (\ref{EL}) with the identity%
\begin{equation}
\frac{d\mathrm{L}\left(  r^{\prime}\text{, }r\right)  }{d\theta}%
=\frac{\partial\mathrm{L}\left(  r^{\prime}\text{, }r\right)  }{\partial
r^{\prime}}r^{\prime\prime}+\frac{\partial\mathrm{L}\left(  r^{\prime}\text{,
}r\right)  }{\partial r}r^{\prime}\text{,}%
\end{equation}
we get%
\begin{equation}
\frac{d\mathrm{L}\left(  r^{\prime}\text{, }r\right)  }{d\theta}=\frac
{d}{d\theta}\left(  r^{\prime}\frac{\partial\mathrm{L}\left(  r^{\prime
}\text{, }r\right)  }{\partial r^{\prime}}\right)  \text{.} \label{be}%
\end{equation}
Eq. (\ref{be}) finally leads to the Beltrami identity in Eq. (\ref{Beltrami}).
Using Eqs. (\ref{s2}) and (\ref{Beltrami}), we get \ \ \
\begin{equation}
\frac{1}{\sqrt{1+\frac{r^{\prime2}}{1-r^{2}}}}=\text{\textrm{const}.}%
\equiv\mathrm{c}_{\mathrm{S}}\text{,} \label{c}%
\end{equation}
that is,%
\begin{equation}
\frac{r^{\prime2}}{1-r^{2}}=\text{\textrm{const}.}\equiv\mathrm{k}%
\overset{\text{def}}{=}\frac{1-\mathrm{c}_{\mathrm{S}}^{2}}{\mathrm{c}%
_{\mathrm{S}}^{2}}\text{,} \label{be2}%
\end{equation}
Integrating Eq. (\ref{be2}) and imposing the boundary conditions $r\left(
\theta_{i}\right)  =r_{i}$ and $r\left(  \theta_{f}\right)  =r_{f}$ with
$\theta_{i}=0$, we finally get the geodesic path%
\begin{equation}
r_{\mathrm{Sj\ddot{o}qvist}}\left(  \theta\right)  =\sin\left[  \sin
^{-1}\left(  r_{i}\right)  +\frac{\sin^{-1}\left(  r_{f}\right)  -\sin
^{-1}\left(  r_{i}\right)  }{\theta_{f}}\theta\right]  \text{.} \label{rerik}%
\end{equation}
Evaluating Eq. (\ref{c}) at $\theta_{i}=0$ and using Eq. (\ref{rerik}), we
note that the constant $\mathrm{c}_{\mathrm{S}}$ can be expressed in terms of
$r_{i}$, $r_{f}$, and $\theta_{f}$ as%
\begin{equation}
\mathrm{c}_{\mathrm{S}}=\mathrm{c}_{\mathrm{S}}\left(  r_{i}\text{, }%
r_{f}\text{, }\theta_{f}\right)  \overset{\text{def}}{=}\frac{\theta_{f}%
}{\sqrt{\theta_{f}^{2}+\left[  \sin^{-1}\left(  r_{f}\right)  -\sin
^{-1}\left(  r_{i}\right)  \right]  ^{2}}}\text{,}%
\end{equation}
with $0\leq\mathrm{c}_{\mathrm{S}}\leq1$, as correctly expected. For
completeness, we point out that in terms of initial conditions $r\left(
\theta_{i}\right)  =r_{i}$ and $r^{\prime}\left(  \theta_{i}\right)
=r_{i}^{\prime}$ with $\theta_{i}=0$, $r_{\mathrm{Sj\ddot{o}qvist}}\left(
\theta\right)  $ in Eq. (\ref{rerik}) can be recast as%
\begin{equation}
r_{\mathrm{Sj\ddot{o}qvist}}\left(  \theta\right)  =\sin\left[  \sin
^{-1}\left(  r_{i}\right)  +\frac{r_{i}^{\prime}}{\sqrt{1-r_{i}^{2}}}%
\theta\right]  \text{,} \label{rerik1}%
\end{equation}
where we used $\cos\left[  \sin^{-1}\left(  r_{i}\right)  \right]
=\sqrt{1-r_{i}^{2}}$ in the raw solution of the form $r\left(  \theta\right)
=\sin\left(  c_{1}\theta+c_{2}\right)  $ with real integrations constants
$c_{1}$ and $c_{2}$. As a final remark, observe from Eq. (\ref{rerik}) that
$\theta$ plays the role of the affine parameter that characterizes the points
on the curve of minimal length connecting the initial and final points in the
Bloch ball.

\subsection{The canonical $\eta$-affine parametrization}

To find the geodesic paths in the Bloch ball suing the Sj\"{o}qvist metric and
parametrized in terms of the \textquotedblleft canonical\textquotedblright%
\ affine parameter $\eta$, we need to integrate Eq. (\ref{GE2}). We note from
Eq. (\ref{GE2}) that the differential equations for the angular and radial
motion are not coupled. In particular, the angular motion is identical to the
one that emerges when using the Fubini-Study metric. Therefore, we refer to
Appendix A for the characterization of the angular motion. Let us focus here
on the radial motion specified by the relation%
\begin{equation}
\ddot{r}+\frac{r}{1-r^{2}}\dot{r}^{2}=0\text{,} \label{star}%
\end{equation}
with $\dot{r}\overset{\text{def}}{=}dr/d\eta$. Assuming initial conditions
given by $r\left(  \eta_{i}\right)  =r_{i}$ and $\dot{r}\left(  \eta
_{i}\right)  =\dot{r}_{i}$ with $\eta_{i}=0$, integration of Eq. (\ref{star})
yields%
\begin{equation}
r_{\mathrm{Sj\ddot{o}qvist}}\left(  \eta\right)  =\sin\left[  \sin^{-1}\left(
r_{i}\right)  +\frac{\dot{r}_{i}}{\sqrt{1-r_{i}^{2}}}\eta\right]  \text{.}
\label{star1}%
\end{equation}
Eq. (\ref{star1}) represents the radial geodesic path parametrized in terms of
the affine parameter $\eta$, the time coordinate, related to the proper length
$ds_{\mathrm{Sj\ddot{o}qvist}}$. Note that the speed of evolution along
geodesic paths is constant and equals,%
\begin{equation}
v_{\mathrm{Sj\ddot{o}qvist}}\overset{\text{def}}{=}\left(  1/2\right)  \left[
\left(  1-r^{2}\right)  ^{-1}\dot{r}^{2}+\dot{\theta}^{2}+\sin^{2}\left(
\theta\right)  \dot{\varphi}^{2}\right]  ^{1/2}\text{.} \label{geospeed}%
\end{equation}
The constancy of $v_{\mathrm{Sj\ddot{o}qvist}}$ in Eq. (\ref{geospeed}) can be
verified by exploiting the constancy of $v_{\mathrm{FS}}$ along with the
constancy of $\left(  1-r^{2}\right)  ^{-1}\dot{r}^{2}$ by means of Eq.
(\ref{star1}). Interestingly, assuming $\varphi=$\textrm{const.}, the geodesic
equation $\theta\left(  \eta\right)  $ becomes $\ddot{\theta}=0$. Therefore,
assuming $\theta\left(  \eta_{i}\right)  =0$ and $\theta\left(  \eta
_{f}\right)  =\theta_{f}$, we get $\theta\left(  \eta\right)  =\left(
\theta_{f}/\eta_{f}\right)  \eta$. Then, considering an affine change of
variables defined by $\eta\rightarrow\theta=\theta\left(  \eta\right)
\overset{\text{def}}{=}\left(  \theta_{f}/\eta_{f}\right)  \eta$, Eq.
(\ref{star}) becomes%
\begin{equation}
r^{\prime\prime}+\frac{r}{1-r^{2}}r^{\prime2}=0\text{,} \label{star2}%
\end{equation}
with $r^{\prime}\overset{\text{def}}{=}dr/d\eta$. Finally, assuming $r\left(
\theta_{i}\right)  =r_{i}$ and $r\left(  \theta_{f}\right)  =r_{f}$,
integration of Eq. (\ref{star2}) yields exactly $r_{\mathrm{Sj\ddot{o}qvist}%
}\left(  \theta\right)  $ in Eq. (\ref{rerik}).

\section{Geodesic paths in the Bloch ball with the Bures metric}

In this Appendix, we study the geodesic paths in the Bloch ball using the
Bures metric.

\subsection{The Sj\"{o}qvist-like $\theta$-affine parametrization}

Following the line of reasoning used in the Sj\"{o}qvist affine
parametrization case in Appendix C, we use spherical coordinates $\left(
r\text{, }\theta\text{, }\varphi\right)  $ and keep $\varphi=$\textrm{const}.
Then, geodesics are obtained by minimizing $\int$ $ds_{\mathrm{Bures}}$ over
all curves connecting points $\left(  r_{i}\text{, }\theta_{i}\right)  $ and
$\left(  r_{f}\text{, }\theta_{f}\right)  $. More specifically, one arrives at
the curve $\left[  \theta_{i}\text{, }\theta_{f}\right]  \ni\theta\mapsto
r_{\mathrm{Bures}}\left(  \theta\right)  \in\left[  0\text{, }1\right]  $ that
minimizes the length $\mathcal{L}_{\mathrm{Bures}}\left(  \theta_{i}\text{,
}\theta_{f}\right)  $ defined as%
\begin{equation}
\mathcal{L}_{\mathrm{Bures}}\left(  \theta_{i}\text{, }\theta_{f}\right)
\overset{\text{def}}{=}\int_{s_{i}}^{s_{f}}\sqrt{ds_{\mathrm{Bures}}^{2}%
}=\frac{1}{2}\int_{\theta_{i}}^{\theta_{f}}\mathrm{L}\left(  r^{\prime}\text{,
}r\text{, }\theta\right)  d\theta\text{.} \label{s1a}%
\end{equation}
In Eq. (\ref{s1a}), $r^{\prime}\overset{\text{def}}{=}dr/d\theta$ and
$\mathrm{L}\left(  r^{\prime}\text{, }r\text{, }\theta\right)  $ is the
Lagrangian-like function given by%
\begin{equation}
\mathrm{L}\left(  r^{\prime}\text{, }r\text{, }\theta\right)
\overset{\text{def}}{=}\sqrt{r^{2}+\frac{r^{\prime2}}{1-r^{2}}}\text{.}
\label{s2a}%
\end{equation}
From Eq. (\ref{s2a}), we follow step-by-step the analysis carried out in the
Sj\"{o}qvist metric case and arrive at the following analog of Eq. (\ref{c})
\ \ \
\begin{equation}
\frac{r^{2}}{\sqrt{r^{2}+\frac{r^{\prime2}}{1-r^{2}}}}=\text{\textrm{const}%
.}\equiv\mathrm{c}_{\mathrm{B}}\text{.} \label{ca}%
\end{equation}
Manipulating Eq. (\ref{ca}) and imposing the boundary conditions $r\left(
\theta_{i}\right)  =r_{i}$ and $r\left(  \theta_{f}\right)  =r_{f}$ with
$\theta_{i}=0$, we obtain%
\begin{equation}
\int_{r_{i}}^{r_{f}}\frac{dr}{\sqrt{r^{2}\left(  \frac{r^{2}}{\mathrm{c}%
_{\mathrm{B}}^{2}}-1\right)  \left(  1-r^{2}\right)  }}=\int_{\theta_{i}%
}^{\theta_{f}}d\theta\text{,} \label{ca1}%
\end{equation}
with $0<\mathrm{c}_{\mathrm{B}}\leq r\leq1$. For notational simplicity, let us
put $\mathrm{a}_{\mathrm{B}}\overset{\text{def}}{=}1/\mathrm{c}_{\mathrm{B}%
}^{2}>1$. Then, integration of Eq. (\ref{ca1}) by use of Mathematica gives%
\begin{equation}
\mathrm{I}\left(  r\right)  \overset{\text{def}}{=}\int\frac{dr}{\sqrt
{r^{2}\left(  \frac{r^{2}}{\mathrm{c}_{\mathrm{B}}^{2}}-1\right)  \left(
1-r^{2}\right)  }}=\frac{r\sqrt{r^{2}-1}\sqrt{\mathrm{a}_{\mathrm{B}}r^{2}-1}%
}{\sqrt{-r^{2}\left(  r^{2}-1\right)  \left(  \mathrm{a}_{\mathrm{B}}%
r^{2}-1\right)  }}\tanh^{-1}\left(  \frac{\sqrt{r^{2}-1}}{\sqrt{\mathrm{a}%
_{\mathrm{B}}r^{2}-1}}\right)  +\text{\textrm{const}.} \label{ca2}%
\end{equation}
Manipulation of Eq. (\ref{ca2}) yields%
\begin{equation}
\mathrm{I}\left(  r\right)  =i\tanh^{-1}\left(  i\frac{\sqrt{1-r^{2}}}%
{\sqrt{\mathrm{a}_{\mathrm{B}}r^{2}-1}}\right)  +\text{\textrm{const}.,}%
\end{equation}
that is,%
\begin{equation}
\mathrm{I}\left(  r\right)  =-\arctan\left(  \frac{\sqrt{1-r^{2}}}%
{\sqrt{\mathrm{a}_{\mathrm{B}}r^{2}-1}}\right)  +\text{\textrm{const}.}
\label{ca3}%
\end{equation}
Substituting Eq. (\ref{ca3}) into Eq. (\ref{ca2}), we finally get that the
radial geodesic path in the Bures case is given by%
\begin{equation}
r_{\mathrm{Bures}}\left(  \theta\right)  =\left\{  \frac{1+\tan^{2}\left[
\mathcal{A}-\left(  \theta-\theta_{i}\right)  \right]  }{1+\mathrm{a}%
_{\mathrm{B}}\tan^{2}\left[  \mathcal{A}-\left(  \theta-\theta_{i}\right)
\right]  }\right\}  ^{1/2}\text{,} \label{ca4}%
\end{equation}
where the constant $\mathcal{A}$ in Eq. (\ref{ca4}) is defined as%
\begin{equation}
\mathcal{A}=\mathcal{A}\left(  r_{i}\text{, }\mathrm{a}_{\mathrm{B}}\right)
\overset{\text{def}}{=}\arctan\left(  \frac{\sqrt{1-r_{i}^{2}}}{\sqrt
{\mathrm{a}_{\mathrm{B}}r_{i}^{2}-1}}\right)  \text{.}%
\end{equation}
For consistency check, we note that we correctly obtain $r_{\mathrm{Bures}%
}\left(  \theta_{i}\right)  =r_{i}$ with $r_{\mathrm{Bures}}\left(
\theta\right)  $ in Eq. (\ref{ca4}). Moreover, $0\leq r_{\mathrm{Bures}%
}\left(  \theta\right)  \leq1$ since $\mathrm{a}_{\mathrm{B}}>1$ in Eq.
(\ref{ca4}). Clearly, Eq. (\ref{ca4}) with $\theta_{i}$ set equal to zero
should be compared, for a given pair of initial conditions $r_{i}$ and
$r_{i}^{\prime}$, with its corresponding analog in the framework of
Sj\"{o}qvist metric (that is, Eq. (\ref{rerik1})). Such comparison can be
carried out once we express $\mathrm{a}_{\mathrm{B}}$ and $\mathcal{A}$ in Eq.
(\ref{ca4}) in terms of $r_{i}$ and $r_{i}^{\prime}$. After some algebra, it
happens that%
\begin{equation}
\mathrm{a}_{\mathrm{B}}=\mathrm{a}_{\mathrm{B}}\left(  r_{i}\text{, }%
r_{i}^{\prime}\right)  \overset{\text{def}}{=}\frac{1}{r_{i}^{2}}+\frac
{r_{i}^{\prime2}}{r_{i}^{4}\left(  1-r_{i}^{2}\right)  }\text{, and
}\mathcal{A}=\mathcal{A}\left(  r_{i}\text{, }r_{i}^{\prime}\right)
\overset{\text{def}}{=}\tan^{-1}\left[  \frac{r_{i}}{r_{i}^{\prime}}\left(
1-r_{i}^{2}\right)  \right]  \text{.} \label{cac1}%
\end{equation}
For completeness, we point out that $\mathrm{a}_{\mathrm{B}}$ and
$\mathcal{A}$ can only be implicitly expressed in terms of the two boundary
conditions $r_{i}=r\left(  \theta_{i}\right)  $ and $r_{f}=r\left(  \theta
_{f}\right)  $ via the two relations%
\begin{equation}
r_{i}^{2}=\frac{1+\tan^{2}\mathcal{A}}{1+\mathrm{a}_{\mathrm{B}}\tan
^{2}\mathcal{A}}\text{, and }r_{f}^{2}=\frac{1+\tan^{2}\left(  \mathcal{A}%
-\theta_{f}\right)  }{1+\mathrm{a}_{\mathrm{B}}\tan^{2}\left(  \mathcal{A}%
-\theta_{f}\right)  }\text{.}%
\end{equation}
Finally, setting $\theta_{i}=0$ and using Eq. (\ref{cac1}), $r_{\mathrm{Bures}%
}\left(  \theta\right)  $ in Eq. (\ref{ca4}) can be formally compared with
$r_{\mathrm{Sj\ddot{o}qvist}}\left(  \theta\right)  $ in Eq. (\ref{rerik1}).

\subsection{The canonical $\eta$-affine parametrization}

From Eq. (\ref{g12}), the only nonvanishing Christoffel connection
coefficients are%
\begin{align}
\Gamma_{11}^{1}  &  =\frac{r}{1-r^{2}}\text{, }\Gamma_{22}^{1}=-\left(
1-r^{2}\right)  r\text{, }\Gamma_{33}^{1}=-\left(  1-r^{2}\right)  r\sin
^{2}\left(  \theta\right)  \text{,}\nonumber\\
& \nonumber\\
\Gamma_{12}^{2}  &  =\Gamma_{21}^{2}=\frac{1}{r}\text{, }\Gamma_{33}^{2}%
=-\sin\left(  \theta\right)  \cos\left(  \theta\right)  \text{,}\\
& \nonumber\\
\Gamma_{13}^{3}  &  =\Gamma_{31}^{3}=\frac{1}{r}\text{, }\Gamma_{23}%
^{3}=\Gamma_{32}^{3}=\frac{\cos\left(  \theta\right)  }{\sin\left(
\theta\right)  }\text{.}\nonumber
\end{align}
Therefore, geodesics satisfy the geodesic equations in Eq. (\ref{ge})
described in terms of a system of three coupled second order nonlinear ODEs,%
\begin{equation}
\left\{
\begin{array}
[c]{c}%
\ddot{r}+\frac{r}{1-r^{2}}\dot{r}^{2}-\left(  1-r^{2}\right)  r\left[
\dot{\theta}^{2}+\sin^{2}\left(  \theta\right)  \dot{\varphi}^{2}\right]  =0\\
\ddot{\theta}+\frac{2}{r}\dot{r}\dot{\theta}-\sin\left(  \theta\right)
\cos\left(  \theta\right)  \dot{\varphi}^{2}=0\\
\ddot{\varphi}+\frac{2}{r}\dot{r}\dot{\varphi}+2\frac{\cos\left(
\theta\right)  }{\sin\left(  \theta\right)  }\dot{\theta}\dot{\varphi}=0
\end{array}
\right.  \text{,} \label{GE3}%
\end{equation}
where $\dot{r}\overset{\text{def}}{=}dr/d\eta$ with $\eta$ being an affine
parameter. We note from Eqs. (\ref{GE2}) and (\ref{GE3}) that, unlike what
happens with the Sj\"{o}qvist metric, when using the Bures metric in the Bloch
ball, the radial and angular evolutions are coupled. Integration of the system
in Eq. (\ref{GE3}) is rather complicated and, in what follows, we limit our
attention to geodesic paths with constant $\varphi$. In this case, Eq.
(\ref{GE3}) reduces to%
\begin{equation}
\left\{
\begin{array}
[c]{c}%
\ddot{r}+\frac{r}{1-r^{2}}\dot{r}^{2}-\left(  1-r^{2}\right)  r\dot{\theta
}^{2}=0\\
\ddot{\theta}+\frac{2}{r}\dot{r}\dot{\theta}=0
\end{array}
\right.  \text{.} \label{GE3b}%
\end{equation}
Manipulating the second relation in Eq. (\ref{GE3b}), we note that $r^{2}%
\dot{\theta}=\mathrm{const.}$Then, given our knowledge of $r=r\left(
\theta\right)  $ in Eq. (\ref{ca4}), we get%
\begin{equation}
\int_{\theta_{i}}^{\theta}r^{2}\left(  \theta\right)  d\theta=r_{i}^{2}%
\dot{\theta}_{i}\eta\text{,} \label{cac11}%
\end{equation}
where $\eta_{i}$ is assumed to be equal to zero. Using Eq. (\ref{ca4}),
integration of Eq. (\ref{cac11}) with Mathematica along with some algebraic
manipulations yield $\theta=\theta\left(  \eta\right)  $ as%
\begin{equation}
\theta_{\mathrm{Bures}}\left(  \eta\right)  =\theta_{i}+\mathcal{A}-\tan
^{-1}\left\{  \frac{1}{\sqrt{\mathrm{a}_{\mathrm{B}}}}\tan\left[  \tan
^{-1}\left(  \sqrt{\mathrm{a}_{\mathrm{B}}}\tan\mathcal{A}\right)
-\sqrt{\mathrm{a}_{\mathrm{B}}}r_{i}^{2}\dot{\theta}_{i}\eta\right]  \right\}
\text{.} \label{tetab}%
\end{equation}
Observe that $\theta_{\mathrm{Bures}}\left(  \eta\right)  $ is a bounded
function for any $\eta\geq0$. For consistency, note that we correctly obtain
from Eq. (\ref{tetab}) that $\theta_{\mathrm{Bures}}\left(  \eta_{i}\right)
=\theta_{i}$ and $\dot{\theta}_{\mathrm{Bures}}\left(  \eta_{i}\right)
=\dot{\theta}_{i}$. Finally, using Eq. (\ref{tetab}) and recalling that
$r^{2}\dot{\theta}=r_{i}^{2}\dot{\theta}_{i}$, we get $r=r\left(  \eta\right)
$ as%
\begin{equation}
r_{\mathrm{Bures}}\left(  \eta\right)  =\left(  \frac{\mathrm{a}_{\mathrm{B}%
}+\tan^{2}\left[  \tan^{-1}\left(  \sqrt{\mathrm{a}_{\mathrm{B}}}%
\tan\mathcal{A}\right)  -\sqrt{\mathrm{a}_{\mathrm{B}}}r_{i}^{2}\dot{\theta
}_{i}\eta\right]  }{\mathrm{a}_{\mathrm{B}}+\mathrm{a}_{\mathrm{B}}\tan
^{2}\left[  \tan^{-1}\left(  \sqrt{\mathrm{a}_{\mathrm{B}}}\tan\mathcal{A}%
\right)  -\sqrt{\mathrm{a}_{\mathrm{B}}}r_{i}^{2}\dot{\theta}_{i}\eta\right]
}\right)  ^{1/2}\text{.} \label{rbures}%
\end{equation}
Note that, $0\leq r_{\mathrm{Bures}}\left(  \eta\right)  \leq1$ for any
$\eta\geq0$ since $\mathrm{a}_{\mathrm{B}}>1$ in Eq. (\ref{rbures}). As a side
remark, note that the speed of evolution along these geodesic paths with
$\varphi$-fixed is constant and equals,%
\begin{equation}
v_{\mathrm{Bures}}\overset{\text{def}}{=}\left(  1/2\right)  \left[  \left(
1-r^{2}\right)  ^{-1}\dot{r}^{2}+r^{2}\dot{\theta}^{2}\right]  ^{1/2}\text{.}
\label{speed2}%
\end{equation}
The constancy of $v_{\mathrm{Bures}}$ in Eq. (\ref{speed2}) is a consequence
of Eq. (\ref{ca}) along with the constancy of $r^{2}\dot{\theta}$. As a
consistency check, observe that Eq. (\ref{rbures}) correctly leads to
$r_{\mathrm{Bures}}\left(  \eta_{i}\right)  =r_{i}$ and $r_{\mathrm{Bures}%
}^{2}\left(  \eta\right)  \dot{\theta}_{\mathrm{Bures}}\left(  \eta\right)
=r_{i}^{2}\dot{\theta}_{i}=\mathrm{const.}$ Observe that $r_{\mathrm{Bures}%
}\left(  \eta\right)  $ in Eq. (\ref{rbures}) is the analog of
$r_{\mathrm{Sj\ddot{o}qvist}}\left(  \eta\right)  $ in Eq. (\ref{star1}). It
is evident from Eq. (\ref{rbures}) that the radial variable $r_{\mathrm{Bures}%
}\left(  \eta\right)  $ depends on the angular motion via $\dot{\theta}_{i}$
and $r_{i}^{\prime}=(dr/d\theta)_{\theta=\theta_{i}}$. This latter quantity
enters the expressions of $\mathrm{a}_{\mathrm{B}}$ and $\mathcal{A}$ as
presented in Eq. (\ref{cac1}).

In what follows, we briefly compare features that appear in the Bures and
Sj\"{o}qvist cases. Using Eq. (\ref{ca}) and exploiting the constancy of
$r^{2}\dot{\theta}$, we have%
\begin{equation}
\mathcal{L}_{\mathrm{Bures}}\left(  \theta_{f}\right)  =\frac{1}{2}\sqrt
{r_{i}^{2}+\frac{r_{i}^{\prime2}}{1-r_{i}^{2}}}\dot{\theta}_{i}\eta_{f}%
\leq\mathcal{L}_{\mathrm{Sj\ddot{o}qvist}}\left(  \theta_{f}\right)  \text{,}
\label{amo2}%
\end{equation}
with $\mathcal{L}_{\mathrm{Sj\ddot{o}qvist}}\left(  \theta_{f}\right)  $ in
Eq. (\ref{sl1}). Moreover, using Eqs. (\ref{A28}), (\ref{star1}),
(\ref{tetab}), and (\ref{rbures}) and noting that $V_{\mathrm{Sj\ddot{o}%
qvist}}^{\left(  \text{accessible}\right)  }=\pi^{2}/4=2V_{\mathrm{Bures}%
}^{\left(  \text{accessible}\right)  }$, we get after some algebra that for
sufficiently large values of $\eta$ and fixing $\varphi$,%
\begin{equation}
\frac{V_{\mathrm{Bures}\text{, }\varphi=\mathrm{const.}}^{\left(
\text{explored}\right)  }\left(  \eta\right)  }{V_{\mathrm{Bures}}^{\left(
\text{accessible}\right)  }}\leq\frac{V_{\mathrm{Sj\ddot{o}qvist}\text{,
}\varphi=\mathrm{const.}}^{\left(  \text{explored}\right)  }\left(
\eta\right)  }{V_{\mathrm{Sj\ddot{o}qvist}}^{\left(  \text{accessible}\right)
}}\text{.} \label{boys}%
\end{equation}
In particular, the qualitative behavior of the explored volumes in Eq.
(\ref{boys}) is given by
\begin{equation}
V_{\mathrm{Sj\ddot{o}qvist}\text{, }\varphi=\mathrm{const}}\left(
\eta\right)  \sim\eta\cos^{-1}\left[  \sin\left(  \eta\right)  \right]
\text{, and }V_{\mathrm{Bures}\text{, }\varphi=\mathrm{const}}\left(
\eta\right)  \sim\sqrt{\frac{\tan^{2}\left(  \eta\right)  }{1+\tan^{2}\left(
\eta\right)  }}\tan^{-1}\left(  \eta\right)  \text{.}%
\end{equation}
Since the sequential application of the averaging and asymptotic limit
procedures preserve the ranking in Eq. (\ref{boys}), we expect that the
complexity of the evolution on the Bures manifold is softer than that on the
Sj\"{o}qvist manifold. This is not completely unexpected given that
$\mathcal{L}_{\mathrm{Bures}}\left(  \theta_{f}\right)  \leq\mathcal{L}%
_{\mathrm{Sj\ddot{o}qvist}}\left(  \theta_{f}\right)  $ and, above all, the
presence of a correlational structure in the equations of motion between the
radial and angular directions. Such a structure is absent in the Sj\"{o}qvist
case. Correlational structures do tend to shrink the explored volumes of
regions on the manifold underlying the dynamics and, therefore, tend to weaken
the complexity of the evolution \cite{mancio10,mancio11}. In summary, although
Eq. (\ref{boys}) assumes $\varphi=\mathrm{const.}$, the information geometric
complexity of the evolution of quantum systems in a mixed quantum states seems
to depend on the choice of the metric selected on the underlying manifold.

\section{Curvature of quantum state manifolds}

In this Appendix, we outline some curvature properties of the manifold of pure
states equipped with the Fubini-Study metric along with those of a manifold of
mixed quantum states endowed with the Sj\"{o}qvist and Bures metrics. In
particular, for each case, we report expressions of the tensor metric
components, infinitesimal line elements, Christoffel connection coefficients,
Ricci tensor components, Riemann curvature tensor components, scalar
curvatures and, finally, sectional curvatures.

\subsection{Preliminaries}

Given a metric tensor $g_{\mu\nu}\left(  \xi\right)  $ with corresponding line
element $ds^{2}\overset{\text{def}}{=}$ $g_{\mu\nu}d\xi^{\mu}d\xi^{\nu}$, the
Christoffel connection coefficients are defined as \cite{weinberg},%
\begin{equation}
\Gamma_{\mu\nu}^{\rho}\overset{\text{def}}{=}\frac{1}{2}g^{\rho\sigma}\left(
\partial_{\mu}g_{\sigma\nu}+\partial_{\nu}g_{\mu\sigma}-\partial_{\sigma
}g_{\mu\nu}\right)  \text{,} \label{Christoffel}%
\end{equation}
where $\partial_{\mu}\overset{\text{def}}{=}\partial/\partial\xi^{\mu}$ and
$g^{\rho\sigma}$ are the coefficients of the inverse metric tensor such that
$g^{\rho\sigma}g_{\sigma\beta}\overset{\text{def}}{=}\delta_{\beta}^{\rho}$
with $\delta$ denoting the Kronecker delta symbol. From the expression of the
Christoffel connection coefficients in Eq. (\ref{Christoffel}), the Ricci
tensor and Riemann curvature tensor components can be defined as
\cite{weinberg}%
\begin{equation}
\mathcal{R}_{\mu\nu}\overset{\text{def}}{=}\partial_{\alpha}\Gamma_{\mu\nu
}^{\alpha}-\partial_{\nu}\Gamma_{\mu\alpha}^{\alpha}+\Gamma_{\mu\nu}^{\alpha
}\Gamma_{\alpha\beta}^{\beta}-\Gamma_{\mu\alpha}^{\gamma}\Gamma_{\nu\gamma
}^{\alpha}\text{, } \label{ricci}%
\end{equation}
and,%
\begin{equation}
\mathcal{R}_{\mu\nu\rho}^{\alpha}\overset{\text{def}}{=}\partial_{\nu}%
\Gamma_{\mu\rho}^{\alpha}-\partial_{\rho}\Gamma_{\mu\nu}^{\alpha}%
+\Gamma_{\beta\nu}^{\alpha}\Gamma_{\mu\rho}^{\beta}-\Gamma_{\beta\rho}%
^{\alpha}\Gamma_{\mu\nu}^{\beta}, \label{riemann}%
\end{equation}
respectively. In terms of the quantities in Eqs. (\ref{ricci}) and
(\ref{riemann}), the scalar curvature $\mathcal{R}$ is given by%
\begin{equation}
\mathcal{R}\overset{\text{def}}{=}\mathcal{R}_{\mu\nu}g^{\mu\nu}%
=\mathcal{R}_{\alpha\beta\gamma\delta}g^{\alpha\gamma}g^{\beta\delta}\text{.}
\label{scalar}%
\end{equation}
We remark that the sign of the scalar curvature of a curved manifold is
subject to convention. For instance, following Weinberg's sign convention
\cite{weinberg}, the scalar curvature of a two-sphere of unit radius equals
$-2$. Here, however, we are using the opposite sign convention. In Weinberg's
book, $\left(  \mathcal{R}_{\mu\nu}\right)  _{\mathrm{Weinberg}}%
\overset{\text{def}}{=}-\mathcal{R}_{\mu\nu}$ with $\mathcal{R}_{\mu\nu}$ in
Eq. (\ref{ricci}). Adopting our sign convention, the scalar curvature of a
two-sphere of unit radius equals $+2$. The scalar curvature $\mathcal{R}$ of a
manifold $\mathcal{M}$ in Eq. (\ref{scalar}) can also be recast as the sum of
all sectional curvatures $\mathcal{K}\left(  \hat{e}_{i}\text{, }\hat{e}%
_{j}\right)  $ of planes spanned by pairs $\left\{  \hat{e}_{i}\text{, }%
\hat{e}_{j}\right\}  $ of orthonormal basis elements $\left\{  \hat{e}%
_{k}\right\}  $ with $1\leq k\leq\left\vert T_{p}\mathcal{M}\right\vert $
\cite{lee},%
\begin{equation}
\mathcal{R}=\sum_{i\neq j}\mathcal{K}\left(  \hat{e}_{i}\text{, }\hat{e}%
_{j}\right)  \text{.}%
\end{equation}
The pair $\left\{  \hat{e}_{i}\text{, }\hat{e}_{j}\right\}  $ is a basis for a
$2$-plane $\Pi\subset T_{P}\mathcal{M}$, a two-dimensional subspace of the
tangent space to $\mathcal{M}$ at $P$. The sectional curvature $\mathcal{K}%
\left(  \hat{e}_{i}\text{, }\hat{e}_{j}\right)  $ is defined as
\cite{weinberg},%
\begin{equation}
\mathcal{K}\left(  \hat{e}_{i}\text{, }\hat{e}_{j}\right)  \overset{\text{def}%
}{=}\frac{\mathrm{Riemann}(\hat{e}_{i}\text{, }\hat{e}_{j}\text{, }\hat{e}%
_{j}\text{, }\hat{e}_{i})}{\left\langle \hat{e}_{i}\text{, }\hat{e}%
_{i}\right\rangle \left\langle \hat{e}_{j}\text{, }\hat{e}_{j}\right\rangle
-\left\langle \hat{e}_{i}\text{, }\hat{e}_{j}\right\rangle ^{2}}
\label{sectional}%
\end{equation}
where $\mathrm{Riemann}(a$, $b$, $b$, $a)\overset{\text{def}}{=}%
\mathcal{R}_{\alpha\beta\gamma\delta}a^{\alpha}b^{\beta}b^{\gamma}a^{\delta}$
with $a$, $b$ being two arbitrary vectors on the $2$-plane $\Pi$ spanned by
$\left\{  \hat{e}_{i}\text{, }\hat{e}_{j}\right\}  $ and, finally,
$\left\langle a\text{, }b\right\rangle \overset{\text{def}}{=}g_{\mu\nu}%
a^{\mu}b^{\nu}$. The constancy of the sectional curvatures is related to the
concept of maximally symmetric manifold. Specifically, an isotropic
$n$-dimensional manifold $\mathcal{M}$ is a maximally symmetric manifold with
$n(n+1)/2$ independent Killing vectors where the geometry does not depend on
directions. For a maximally symmetric manifold, the following simplifying
relations hold true among the scalar curvature $\mathcal{R}$, the constant
sectional curvature $\mathcal{K}$, the Ricci tensor components $\mathcal{R}%
_{\alpha\beta}$, and the Riemann curvature tensor components $\mathcal{R}%
_{\alpha\beta\gamma\delta}$ \cite{defelice},%
\begin{equation}
\mathcal{R}=n(n-1)\mathcal{K}\text{, \ }\mathcal{R}_{\alpha\beta
}=(n-1)\mathcal{K}g_{\alpha\beta}\text{, }\mathcal{R}_{\alpha\beta\gamma
\delta}=\frac{\mathcal{R}}{n(n-1)}\text{ }\left(  g_{\beta\delta}%
g_{\alpha\gamma}-g_{\beta\gamma}g_{\alpha\delta}\right)  \text{.}
\label{felice}%
\end{equation}
Isometries play a key role in the characterization of maximally symmetric
manifolds. Recall that an isometry of the metric $g_{\mu\nu}\left(
\xi\right)  $ is a distance-preserving transformation $\xi\rightarrow
\xi^{\prime}$ such that \cite{weinberg},%
\begin{equation}
g_{\mu\nu}\left(  \xi\right)  =\frac{\partial\xi^{\prime\rho}}{\partial
\xi^{\mu}}\frac{\partial\xi^{\prime\sigma}}{\partial\xi^{\nu}}g_{\rho\sigma
}\left(  \xi^{\prime}\right)  \text{.}%
\end{equation}
All the infinitesimal isometries of the metric $g_{\mu\nu}\left(  \xi\right)
$ are determined by the Killing vectors of the metric. Consider an
infinitesimal coordinate transformation $\xi^{\mu}\rightarrow\xi^{\prime\mu
}\overset{\text{def}}{=}\xi^{\mu}+\epsilon k^{\mu}\left(  \xi\right)  $, with
$\left\vert \epsilon\right\vert \ll1$. A vector field $k^{\mu}\left(
\xi\right)  $ is a Killing vector for the metric $g_{\mu\nu}\left(
\xi\right)  $ if it satisfies the so-called Killing condition,%
\begin{equation}
D_{\rho}k_{\sigma}+D_{\sigma}k_{\rho}=0\text{,} \label{killing}%
\end{equation}
where $D_{\rho}k_{\sigma}\overset{\text{def}}{=}\partial_{\rho}k_{\sigma
}-\Gamma_{\sigma\rho}^{\lambda}k_{\lambda}$ and $\partial_{\rho}%
\overset{\text{def}}{=}\partial/\partial\xi^{\rho}$. For completeness, we
point out that what really determines the infinitesimal isometries of a metric
$g_{\mu\nu}\left(  \xi\right)  $ is the space of vector fields spanned by the
Killing vectors since any linear combination of Killing vectors with constant
coefficients is a Killing vector. In general, it is highly nontrivial solving
the Killing conditions in Eq. (\ref{killing}).

\subsection{Type of manifold}

\subsubsection{Manifold equipped with the Fubini-Study metric}

In the case of the two-dimensional manifold of pure states equipped with the
Fubini-Study metric $g_{\mu\nu}^{\mathrm{FS}}\left(  \xi\right)  $ with
$\xi\overset{\text{def}}{=}\left(  \xi^{1}\text{, }\xi^{2}\right)  =\left(
\theta\text{, }\varphi\right)  $, the infinitesimal line element is given by
$ds_{\mathrm{FS}}^{2}\overset{\text{def}}{=}(1/4)\left[  d\theta^{2}+\sin
^{2}\left(  \theta\right)  d\varphi^{2}\right]  $. In this case, the nonzero
Christoffel connection coefficients in Eq. (\ref{Christoffel}) are given by%
\begin{equation}
\Gamma_{22}^{1}=-\sin\left(  \theta\right)  \cos\left(  \theta\right)  \text{,
and }\Gamma_{12}^{2}=\frac{\cos\left(  \theta\right)  }{\sin\left(
\theta\right)  }\text{.} \label{x1}%
\end{equation}
Furthermore, using Eq. (\ref{x1}), the nonzero Ricci and Riemann curvature
tensor components in Eqs. (\ref{ricci}) and (\ref{riemann}) are%
\begin{equation}
\mathcal{R}_{11}=1\text{, }\mathcal{R}_{22}=\sin^{2}\left(  \theta\right)
\text{,}%
\end{equation}
and,%
\begin{equation}
\mathcal{R}_{1212}=\frac{1}{4}\sin^{2}\left(  \theta\right)  \text{, }
\label{riemmanfubini}%
\end{equation}
respectively. For completeness, we point out that exploiting the symmetry
properties of the Riemann curvature tensor, we also have $\mathcal{R}%
_{1221}=\mathcal{R}_{2112}=-\mathcal{R}_{1212}$ and $\mathcal{R}%
_{2121}=\mathcal{R}_{1212}$. To calculate the sectional curvatures
$\mathcal{K}\left(  \hat{e}_{i}\text{, }\hat{e}_{j}\right)  $ in Eq.
(\ref{sectional}), we note that the unit tangent vectors $\left\{  \hat{e}%
_{r}\text{, }\hat{e}_{\theta}\text{, }\hat{e}_{\varphi}\right\}  $ in
spherical coordinates are given by%
\begin{align}
\hat{e}_{r}\overset{\text{def}}{=}\frac{\partial_{r}\vec{r}}{\left\Vert
\partial_{r}\vec{r}\right\Vert }  &  =\sin\left(  \theta\right)  \cos\left(
\varphi\right)  \hat{x}+\sin\left(  \theta\right)  \sin\left(  \varphi\right)
\hat{y}+\cos\left(  \theta\right)  \hat{z}\text{,}\nonumber\\
\hat{e}_{\theta}\overset{\text{def}}{=}\frac{\partial_{\theta}\vec{r}%
}{\left\Vert \partial_{\theta}\vec{r}\right\Vert }  &  =\cos\left(
\theta\right)  \cos\left(  \varphi\right)  \hat{x}+\cos\left(  \theta\right)
\sin\left(  \varphi\right)  \hat{y}-\sin\left(  \theta\right)  \hat{z}%
\text{,}\nonumber\\
\hat{e}_{\varphi}\overset{\text{def}}{=}\frac{\partial_{\varphi}\vec{r}%
}{\left\Vert \partial_{\varphi}\vec{r}\right\Vert }  &  =-\sin\left(
\varphi\right)  \hat{x}+\cos\left(  \varphi\right)  \hat{y}\text{,}
\label{unit}%
\end{align}
where $\vec{r}\overset{\text{def}}{=}r\sin\left(  \theta\right)  \cos\left(
\varphi\right)  \hat{x}+r\sin\left(  \theta\right)  \sin\left(  \varphi
\right)  \hat{y}+r\cos\left(  \theta\right)  \hat{z}$, and $\left\Vert
\cdot\right\Vert $ denotes the usual Euclidean norm. In the Fubini-Study case,
we have that $ds_{\mathrm{FS}}^{2}=\mathbf{ds}_{\mathrm{FS}}\mathbf{\cdot
ds}_{\mathrm{FS}}$ with the infinitesimal vector element $\mathbf{ds}%
_{\mathrm{FS}}$ given by%
\begin{equation}
\mathbf{ds}_{\mathrm{FS}}\overset{\text{def}}{=}\frac{1}{2}\hat{e}_{\theta
}d\theta+\frac{1}{2}\sin\left(  \theta\right)  \hat{e}_{\varphi}%
d\varphi\text{,} \label{minghi1}%
\end{equation}
where $1/2$ and $\left(  1/2\right)  \sin\left(  \theta\right)  $ in Eq.
(\ref{minghi1}) denote the so-called scale factors of the metric \cite{boas}.
Then, inserting Eqs. (\ref{unit}) and (\ref{riemmanfubini}) into Eq.
(\ref{sectional}), we find%
\begin{equation}
\mathcal{K}_{\mathrm{FS}}\left(  \hat{e}_{\theta}\text{, }\hat{e}_{\varphi
}\right)  =\mathcal{K}_{\mathrm{FS}}\left(  \hat{e}_{\varphi}\text{, }\hat
{e}_{\theta}\right)  =4\text{.} \label{secco1}%
\end{equation}
Thus, from Eq. (\ref{secco1}), we conclude that the manifold of pure states
equipped with the Fubini-Study metric is an isotropic manifold of constant
(positive) sectional curvature with (positive) constant Ricci curvature
$\mathcal{R}_{\mathrm{FS}}=8$. As a final remark, we emphasize that for a
two-sphere with metric $4ds_{\mathrm{FS}}^{2}\overset{\text{def}}{=}%
d\theta^{2}+\sin^{2}\left(  \theta\right)  d\varphi^{2}$, Killing vectors can
be found \cite{press}%
\begin{equation}
k_{1}\overset{\text{def}}{=}L_{x}/i\hslash=\sin\left(  \varphi\right)
\partial_{\theta}+\cot\left(  \theta\right)  \cos\left(  \varphi\right)
\partial_{\varphi}\text{, }k_{2}\overset{\text{def}}{=}L_{y}/i\hslash
=-\cos\left(  \varphi\right)  \partial_{\theta}+\cot\left(  \theta\right)
\sin\left(  \varphi\right)  \partial_{\varphi}\text{, }k_{3}%
\overset{\text{def}}{=}L_{z}/i\hslash=-\partial_{\varphi}\text{.}%
\end{equation}
Then, the most general Killing vector $k$ is a linear combination of these
three independent Killing vectors $\left\{  k_{1}\text{, }k_{2}\text{, }%
k_{3}\right\}  $. The three vectors describe rotations and are just the
angular momentum operators $\left\{  L_{x}\text{, }L_{y}\text{, }%
L_{z}\right\}  $, the generators of the three-dimensional rotation group
$SO\left(  3\text{; }%
%TCIMACRO{\U{211d} }%
%BeginExpansion
\mathbb{R}
%EndExpansion
\right)  $, expressed in spherical coordinates \cite{sakurai}.

\subsubsection{Manifold equipped with the Sj\"{o}qvist metric}

For the three-dimensional manifold of mixed states equipped with the
Sj\"{o}qvist metric $g_{\mu\nu}^{\mathrm{Sj\ddot{o}qvist}}\left(  \xi\right)
$ with $\xi\overset{\text{def}}{=}\left(  \xi^{1}\text{, }\xi^{2}\text{, }%
\xi^{3}\right)  =\left(  r\text{, }\theta\text{, }\varphi\right)  $, the
infinitesimal line element is $ds_{\mathrm{Sj\ddot{o}qvist}}^{2}%
\overset{\text{def}}{=}(1/4)\left[  \left(  1-r^{2}\right)  ^{-1}%
dr^{2}+d\theta^{2}+\sin^{2}\left(  \theta\right)  d\varphi^{2}\right]  $. The
nonzero Christoffel connection coefficients in Eq. (\ref{Christoffel}) are%
\begin{equation}
\Gamma_{11}^{1}=\frac{r}{1-r^{2}}\text{, }\Gamma_{33}^{2}=-\sin\left(
\theta\right)  \cos\left(  \theta\right)  \text{, and }\Gamma_{23}^{3}%
=\frac{\cos\left(  \theta\right)  }{\sin\left(  \theta\right)  }\text{.}
\label{x2}%
\end{equation}
Moreover, exploiting Eq. (\ref{x2}), the non vanishing Ricci and Riemann
curvature tensor components in Eqs. (\ref{ricci}) and (\ref{riemann}) become%
\begin{equation}
\mathcal{R}_{22}=1\text{, }\mathcal{R}_{33}=\sin^{2}\left(  \theta\right)
\text{,}%
\end{equation}
and,%
\begin{equation}
\mathcal{R}_{2323}=\frac{1}{4}\sin^{2}\left(  \theta\right)  \text{, }
\label{rere}%
\end{equation}
respectively. For completeness, we emphasize that exploiting the symmetry
properties of the Riemann curvature tensor, we also have $\mathcal{R}%
_{2332}=\mathcal{R}_{3223}=-\mathcal{R}_{2323}$ and $\mathcal{R}%
_{3232}=\mathcal{R}_{2323}$. In the Sj\"{o}qvist metric, we have that
$ds_{\mathrm{Sj\ddot{o}qvist}}^{2}=\mathbf{ds}_{\mathrm{Sj\ddot{o}qvist}%
}\mathbf{\cdot ds}_{\mathrm{Sj\ddot{o}qvist}}$ with the infinitesimal vector
element $\mathbf{ds}_{\mathrm{Sj\ddot{o}qvist}}$ defined as%
\begin{equation}
\mathbf{ds}_{\mathrm{Sj\ddot{o}qvist}}\overset{\text{def}}{=}\frac{1}{2}%
\frac{1}{\sqrt{1-r^{2}}}\hat{e}_{r}dr+\frac{1}{2}\hat{e}_{\theta}d\theta
+\frac{1}{2}\sin\left(  \theta\right)  \hat{e}_{\varphi}d\varphi\text{,}
\label{minghi2}%
\end{equation}
where$\left(  1/2\right)  (1-r^{2})^{-1/2}$, $1/2$ and $\left(  1/2\right)
\sin\left(  \theta\right)  $ in Eq. (\ref{minghi2}) denote the so-called scale
factors of the metric. Then, inserting Eqs. (\ref{unit}) and (\ref{rere}) into
Eq. (\ref{sectional}), we find%
\begin{equation}
\mathcal{K}_{\mathrm{Sj\ddot{o}qvist}}\left(  \hat{e}_{\theta}\text{, }\hat
{e}_{\varphi}\right)  =\mathcal{K}_{\mathrm{Sj\ddot{o}qvist}}\left(  \hat
{e}_{\varphi}\text{, }\hat{e}_{\theta}\right)  =4\text{,} \label{secco2}%
\end{equation}
and,%
\begin{equation}
\mathcal{K}_{\mathrm{Sj\ddot{o}qvist}}\left(  \hat{e}_{r}\text{, }\hat
{e}_{\theta}\right)  =\mathcal{K}_{\mathrm{Sj\ddot{o}qvist}}\left(  \hat
{e}_{\theta}\text{, }\hat{e}_{r}\right)  =\mathcal{K}_{\mathrm{Sj\ddot
{o}qvist}}\left(  \hat{e}_{r}\text{, }\hat{e}_{\varphi}\right)  =\mathcal{K}%
_{\mathrm{Sj\ddot{o}qvist}}\left(  \hat{e}_{\varphi}\text{, }\hat{e}%
_{r}\right)  =0\text{.} \label{secco3}%
\end{equation}
Thus, from Eqs. (\ref{secco2}) and (\ref{secco3}), we conclude that the
manifold of mixed states equipped with the Sj\"{o}qvist metric is an
anisotropic manifold of non-constant (positive) sectional curvature with an
overall (positive) constant Ricci curvature $\mathcal{R}_{\mathrm{Sj\ddot
{o}qvist}}=8$.

\subsubsection{Manifold equipped with the Bures metric}

In the case of the three-dimensional manifold of mixed states equipped with
the Bures metric $g_{\mu\nu}^{\mathrm{Bures}}\left(  \xi\right)  $ with
$\xi\overset{\text{def}}{=}\left(  \xi^{1}\text{, }\xi^{2}\text{, }\xi
^{3}\right)  =\left(  r\text{, }\theta\text{, }\varphi\right)  $, the
infinitesimal line element is given by $ds_{\mathrm{Bures}}^{2}%
\overset{\text{def}}{=}(1/4)\left[  \left(  1-r^{2}\right)  ^{-1}dr^{2}%
+r^{2}d\theta^{2}+r^{2}\sin^{2}\left(  \theta\right)  d\varphi^{2}\right]  $.
In this case, the nonzero Christoffel connection coefficients in Eq.
(\ref{Christoffel}) are given by%
\begin{align}
\Gamma_{11}^{1}  &  =\frac{r}{1-r^{2}}\text{, }\Gamma_{22}^{1}=-r\left(
1-r^{2}\right)  \text{, }\Gamma_{33}^{1}=-r\left(  1-r^{2}\right)  \sin
^{2}\left(  \theta\right)  \text{, }\Gamma_{12}^{2}=\frac{1}{r}\text{,}%
\nonumber\\
& \nonumber\\
\Gamma_{33}^{2}  &  =-\sin\left(  \theta\right)  \cos\left(  \theta\right)
\text{, }\Gamma_{13}^{3}=\frac{1}{r}\text{, }\Gamma_{23}^{3}=\frac{\cos\left(
\theta\right)  }{\sin\left(  \theta\right)  }\text{.} \label{x3}%
\end{align}
Furthermore, employing Eq. (\ref{x3}), the nonzero Ricci and Riemann curvature
tensor components in Eqs. (\ref{ricci}) and (\ref{riemann}) are%
\begin{equation}
\mathcal{R}_{11}=\frac{2}{1-r^{2}}\text{, }\mathcal{R}_{22}=2r^{2}\text{,
}\mathcal{R}_{33}=2r^{2}\sin^{2}\left(  \theta\right)  \text{,}%
\end{equation}
and, modulo symmetries of the Riemann curvature tensor,%
\begin{equation}
\mathcal{R}_{1212}=\frac{1}{4}\frac{r^{2}}{1-r^{2}}\text{, }\mathcal{R}%
_{1313}=\frac{1}{4}\frac{r^{2}}{1-r^{2}}\sin^{2}\left(  \theta\right)  \text{,
}\mathcal{R}_{2323}=\frac{1}{4}r^{4}\sin^{2}\left(  \theta\right)  \text{, }
\label{burino}%
\end{equation}
respectively. In the Bures case, we have that $ds_{\mathrm{Bures}}%
^{2}=\mathbf{ds}_{\mathrm{Bures}}\mathbf{\cdot ds}_{\mathrm{Bures}}$ with the
infinitesimal vector element $\mathbf{ds}_{\mathrm{Bures}}$ given by%
\begin{equation}
\mathbf{ds}_{\mathrm{Bures}}\overset{\text{def}}{=}\frac{1}{2}\frac{1}%
{\sqrt{1-r^{2}}}\hat{e}_{r}dr+\frac{r}{2}\hat{e}_{\theta}d\theta+\frac{r}%
{2}\sin\left(  \theta\right)  \hat{e}_{\varphi}d\varphi\text{,}
\label{minghi3}%
\end{equation}
where $\left(  1/2\right)  (1-r^{2})^{-1/2}$, $r/2$ and $\left(  r/2\right)
\sin\left(  \theta\right)  $ in Eq. (\ref{minghi3}) are the scale factors of
the metric. Then, inserting Eqs. (\ref{unit}) and (\ref{burino}) into Eq.
(\ref{sectional}), we find%
\begin{equation}
\mathcal{K}_{\mathrm{Bures}}\left(  \hat{e}_{r}\text{, }\hat{e}_{\theta
}\right)  =\mathcal{K}_{\mathrm{Bures}}\left(  \hat{e}_{\theta}\text{, }%
\hat{e}_{r}\right)  =\mathcal{K}_{\mathrm{Bures}}\left(  \hat{e}_{r}\text{,
}\hat{e}_{\varphi}\right)  =\mathcal{K}_{\mathrm{Bures}}\left(  \hat
{e}_{\varphi}\text{, }\hat{e}_{r}\right)  =\mathcal{K}_{\mathrm{Bures}}\left(
\hat{e}_{\theta}\text{, }\hat{e}_{\varphi}\right)  =\mathcal{K}%
_{\mathrm{Bures}}\left(  \hat{e}_{\varphi}\text{, }\hat{e}_{\theta}\right)
=4\text{.} \label{YO}%
\end{equation}
Thus, from Eq. (\ref{YO}), we conclude that the manifold of mixed states
equipped with the Bures metric is an isotropic manifold of constant (positive)
sectional curvature with (positive) constant Ricci curvature $\mathcal{R}%
_{\mathrm{Bures}}=24$.

\section{Further details on the Sj\"{o}qvist metric}

In this Appendix, we provide some comparative statements between the Bures and
the Sj\"{o}qvist metrics. Furthermore, we briefly present the extension of the
original Sj\"{o}qvist metric to degenerate mixed quantum states.

\subsection{Comparison with the Bures metric}

Following the Morozova-Cencov-Petz theorem as reported in Ref. \cite{karol},
every (Riemannian and monotone) metric in the Bloch ball at a point where the
density matrix is diagonal, $\rho=\left(  1/2\right)  \mathrm{diag}\left(
1+r\text{, }1-r\right)  $, can be expressed as%
\begin{equation}
ds^{2}=\frac{1}{4}\left[  \frac{dr^{2}}{1-r^{2}}+\frac{1}{f\left(  \frac
{1-r}{1+r}\right)  }\frac{r^{2}}{1+r}d\Omega^{2}\right]  \text{,} \label{MCP}%
\end{equation}
with $0<r<1$. In Eq. (\ref{MCP}), $d\Omega^{2}\overset{\text{def}}{=}%
d\theta^{2}+\sin^{2}\left(  \theta\right)  d\varphi^{2}$ is the metric on the
unit $2$-sphere while $f:%
%TCIMACRO{\U{211d} }%
%BeginExpansion
\mathbb{R}
%EndExpansion
_{+}\rightarrow%
%TCIMACRO{\U{211d} }%
%BeginExpansion
\mathbb{R}
%EndExpansion
_{+}$ is the so-called Morozova-Cencov function $f=f\left(  t\right)  $. This
function satisfies three conditions: (i) $f$ is operator monotone; (ii) $f$ is
self inversive with $f\left(  1/t\right)  =f\left(  t\right)  /t$; (iii)
$f\left(  1\right)  =1$. From Eq. (\ref{MCP}), we emphasize that condition
(iii), $f\left(  1\right)  =1\neq0$, serves to avoid a conical singularity in
the metric at the maximally mixed state where $r=0$ (that is, $t=t\left(
r\right)  \overset{\text{def}}{=}\left(  1-r\right)  /(1+r)=1$). For details
on the Morozova-Cencov-Petz theorem and further discussion on the meaning of
conditions (i)-(ii)-(iii), we refer to Refs. \cite{karol,petz99}. For details
on the monotonicity of operator functions, we refer to Refs.
\cite{bhatia97,kwong89,petz96a,furuta08,gibilisco09}. In the case of the Bures
metric,
\begin{equation}
ds_{\mathrm{Bures}}^{2}=\frac{1}{4}\left[  \frac{dr^{2}}{1-r^{2}}+r^{2}%
d\Omega^{2}\right]  \text{.} \label{BuresMCP}%
\end{equation}
From Eqs. (\ref{MCP}) and (\ref{BuresMCP}), we find that $f_{\mathrm{Bures}%
}\left(  t\right)  \overset{\text{def}}{=}\left(  1+t\right)  /2$. Clearly,
$f_{\mathrm{Bures}}\left(  t\right)  $ satisfies conditions (i), (ii), and
(iii). In the case of the Sj\"{o}qvist metric,
\begin{equation}
ds_{\mathrm{Sj\ddot{o}qvist}}^{2}=\frac{1}{4}\left[  \frac{dr^{2}}{1-r^{2}%
}+d\Omega^{2}\right]  \text{.} \label{SjoqvistMCP}%
\end{equation}
From Eqs. (\ref{MCP}) and (\ref{SjoqvistMCP}), we find that
$f_{\mathrm{Sj\ddot{o}qvist}}\left(  t\right)  \overset{\text{def}}{=}\left(
1/2\right)  \left[  \left(  1-t\right)  ^{2}/\left(  1+t\right)  \right]  $.
We observe that although $f_{\mathrm{Sj\ddot{o}qvist}}\left(  t\right)  $ is
self inversive, $f_{\mathrm{Sj\ddot{o}qvist}}\left(  1\right)  =0$. Therefore,
as pointed out in Ref. \cite{erik20}, the Sj\"{o}qvist metric in Eq.
(\ref{SjoqvistMCP}) is singular at the origin of the Bloch ball where $r=0$
(i.e., $t\equiv t\left(  0\right)  =1$) and the angular components of the
metric tensor diverge because $f_{\mathrm{Sj\ddot{o}qvist}}\left(  1\right)
=0$. For this reason, the original Sj\"{o}qvist metric is limited to
non-degenerate mixed quantum states. When considering degenerate quantum
states $\rho$, the Sj\"{o}qvist metric must be generalized as discussed in
Refs. \cite{silva21,silva21B}. We briefly address this point in the next subsection.

\subsection{Extension to the degenerate case}

From Ref. \cite{erik20} and the main text of this paper (see Eq. (\ref{g6})),
we recall that in the $n$-dimensional case $ds_{\mathrm{Sj\ddot{o}qvist}}^{2}$
can be recast as
\begin{equation}
\left(  ds_{\mathrm{Sj\ddot{o}qvist}}^{2}\right)  ^{\left(  \mathrm{non}%
\text{\textrm{-degenerate}}\right)  }=\frac{1}{4}\sum_{k=1}^{n}\frac
{dp_{k}^{2}}{p_{k}}+\sum_{k=1}^{n}p_{k}ds_{k}^{2}\text{,} \label{G7}%
\end{equation}
where $dp_{k}=\dot{p}_{k}dt$, $ds_{k}^{2}\overset{\text{def}}{=}\left\langle
de_{k}|de_{k}\right\rangle -\left\vert \left\langle e_{k}|de_{k}\right\rangle
\right\vert ^{2}$ is the Fubini-Study metric along the pure state $\left\vert
e_{k}\right\rangle $, and $\hat{1}$ being the identity operator on the
$n$-dimensional Hilbert space. Eq. (\ref{G7}) is valid in the non-degenerate
case. Before introducing its extension to the degenerate case, we make a
remark. Both the Bures and the Sj\"{o}qvist metrics can be viewed as the sum
of a classical and a quantum contribution. In both cases, the classical
contribution is the Fisher-Rao metric between two probability distributions.
The quantum contributions, however, differ in general. In the Bures case (see
Eq. (\ref{ag1})), the quantum contribution emerges from the noncommutativity
of the density matrices $\rho$ and $\rho+d\rho$. When $\left[  \rho
+d\rho\text{, }\rho\right]  =0$, the problem becomes classical and the Bures
metric reduces to the classical Fisher-Rao metric. In the Sj\"{o}qvist case
(see Eq. (\ref{G7})), the quantum contribution is the sum of the pure state
Fubini-Study metrics $ds_{k}^{2}$ along the state vectors $\left\{  \left\vert
e_{k}\right\rangle \right\}  $ weighted with their corresponding probability
$\left\{  p_{k}\right\}  $, $\sum_{k}p_{k}ds_{k}^{2}$. Returning to the
extension of $ds_{\mathrm{Sj\ddot{o}qvist}}^{2}$ in Eq. (\ref{G7}) to the case
of degenerate mixed quantum states, following Refs. \cite{silva21,silva21B},
we have%
\begin{equation}
\left(  ds_{\mathrm{Sj\ddot{o}qvist}}^{2}\right)  ^{\left(
\text{\textrm{degenerate}}\right)  }=\frac{1}{4}\sum_{k=1}^{m}r_{k}%
\frac{dp_{k}^{2}}{p_{k}}+\sum_{k=1}^{m}p_{k}\mathrm{tr}\left(  P_{k}%
dP_{k}dP_{k}\right)  \text{,} \label{G8}%
\end{equation}
where $\rho\overset{\text{def}}{=}\sum_{k=1}^{m}p_{k}P_{k}$, $r_{k}%
\overset{\text{def}}{=}\mathrm{tr}\left(  P_{k}\right)  $ is the rank of the
orthogonal projector $P_{k}$, and $r\overset{\text{def}}{=}\sum_{k=1}^{m}%
r_{k}$ is the rank of the state $\rho$. Obviously, unlike what happens in Eq.
(\ref{G7}), in Eq. (\ref{G8}) not all projectors $P_{k}$ are rank-one
operators because of the possible presence of degenerate eigenvalues of $\rho
$. Note that $m\leq l$ with $l$ denoting the cardinality of the set of pure
states that specify the probabilistic mixture (i.e., quantum ensemble) that
defines $\rho\in%
%TCIMACRO{\U{2102} }%
%BeginExpansion
\mathbb{C}
%EndExpansion
^{n\times n}$. Observe also that $l$ can be greater than $n$ when $\rho$ is
non-degenerate. For a complete classification of quantum ensembles yielding a
given density matrix, we refer to \cite{hughston93}. For completeness, we note
that when $P_{k}=\left\vert e_{k}\right\rangle \left\langle e_{k}\right\vert $
and $r_{k}=1$ for any $1\leq k\leq m$, $\mathrm{tr}\left(  P_{k}dP_{k}%
dP_{k}\right)  $ equals $\left\langle de_{k}|de_{k}\right\rangle -\left\vert
\left\langle e_{k}|de_{k}\right\rangle \right\vert ^{2}$ and, in addition, Eq.
(\ref{G8}) reduces to Eq. (\ref{G7}). For details on the derivation of Eq.
(\ref{G8}), we hint to Refs. \cite{silva21,silva21B}.
\end{document}